\documentclass[aps,prxquantum,twocolumn,superscriptaddress,longbibliography]{revtex4-2}
\pdfoutput=1

\usepackage{amsmath,amssymb,amsfonts}
\usepackage{amsthm}
\usepackage{physics}
\usepackage{braket}
\usepackage{bm}
\usepackage{mathtools}
\usepackage{graphicx}
\usepackage{tikz}
\usepackage{float}
\usepackage{caption}
\usepackage{subcaption}
\usepackage{multirow}
\usepackage{booktabs}
\usepackage{algorithmicx}
\usepackage{algpseudocode}
\usepackage{xcolor}
\usepackage{tcolorbox}
\usepackage[english]{babel}
\usepackage[colorlinks=true,linkcolor=blue,citecolor=blue,urlcolor=blue]{hyperref}
\usepackage{enumitem}
\usepackage{xcolor}

\definecolor{criticalred}{HTML}{C0392B}
\definecolor{highorange}{HTML}{E67E22}
\definecolor{mediumpurple}{HTML}{8E44AD}
\definecolor{lowgray}{HTML}{7F8C8D}

\renewcommand{\Re}{\operatorname{Re}}

\newcommand{\wv}[1]{\,{}^{2}\!\langle #1 \rangle_1}
\newcommand{\mH}{\mathcal{H}}
\providecommand{\Log}{\operatorname{Log}}

\begin{document}

\title{Pseudo-entropy in Thouless Pumps and Majorana Topological Phases}

\author{Nabaraj Khatri}
\thanks{These authors contributed equally to this work.}
\affiliation{Tri-Chandra Multiple Campus, Tribhuvan University, Ghantaghar, 44600 Kathmandu, Nepal}
\affiliation{Holographic Himalaya, Lambagar, 44600 Tarakeshwar, Nepal}

\author{Shuvam Chudal}
\thanks{These authors contributed equally to this work.}
\affiliation{St. Xavier's College, Tribhuvan University, Maitighar, 44600 Kathmandu, Nepal}
\affiliation{Holographic Himalaya, Lambagar, 44600 Tarakeshwar, Nepal}

\author{Ashish Khanal}
\email{ak247626@ohio.edu}
\affiliation{Department of Physics and Astronomy, Ohio University, Athens, OH 45701, USA}
\affiliation{Holographic Himalaya, Lambagar, 44600 Tarakeshwar, Nepal}

\begin{abstract}
We study the average excess pseudo entropy, $\Delta S_{12}$, in the Rice--Mele model under Thouless pumping. We compute it for two driving schemes, the control-freak and smooth pumps, under open and periodic boundary conditions. We find that $\Delta S_{12}$ distinguishes the trivial, monomer, and topological stages of the pump, and that it depends on the boundary condition, the driving scheme, and the reference time. We also find that a nonzero staggered potential breaks chiral symmetry and gaps out the SSH critical point, thereby flattening the pseudo-entropy peak at the phase boundary. We also study reference-state dependence in the open Kitaev chain, comparing a fixed even-parity target with five reference families: even free-fermion, pairing-only, vacuum, and two fixed gapped states. We find that the pseudo-entropy response depends on the reference and on the parity of the subsystem size, even when the reference remains gapped. At $\mu=0$, the pairing-only reference gives the exact result $S_{\mathrm{pr}}=\ln 2$ for $L=4m+2$, while the $L=4m$ sequence has analytically determined finite-size corrections. These results show that, in both models, pseudo entropy is a property of the reference--target pair and the subsystem partition, rather than a reference-independent signature of the topological phase.
\end{abstract}

\maketitle

\section{Introduction}\label{intro}
Quantum information-theoretic tools have received significant attention as useful approaches for studying problems across different areas of physics \cite{nielsen2010quantum, Adhikari_2022}. In particular, quantities such as entanglement entropy, pseudo entropy and quantum complexities have been widely used to investigate phase transitions in condensed matter systems, as well as various aspects of black hole physics and holography \cite{Zeng:2015pxf,calabrese2004entanglement,Kharel:2025lek,Mukherjee:2022jac,Mollabashi:2020yie,Mollabashi:2021xsd,PhysRevLett.128.091301,Brahma:2020zpk,Adhikari:2025zoa}. Entanglement-based measures provide a way to characterize quantum correlations and can reveal important features of quantum phases that may not be easily captured by conventional observables \cite{adhikari2026quantuminformationtheoreticanalysisoptical}. 

In previous work, we studied pseudo entropy as a probe of topological phase transitions using the Su–Schrieffer–Heeger (SSH) model \cite{kharel2026pseudoentropytopologicalphases}. In this work, we extend this study to a non-static system by considering Thouless pumping, in particular, the Rice--Mele model. The Rice--Mele model is a simple lattice model that builds on the SSH chain by adding a staggered on-site potential, allowing both the hopping amplitudes and the on-site energies to be tuned independently \cite{PhysRevLett.49.1455}. When these parameters are varied slowly and periodically in time, the model realizes a Thouless pump in which a fixed integer number of particles is transported across the chain in each cycle, regardless of the details of the driving \cite{PhysRevB.27.6083}. We find that pseudo entropy successfully tracks the different stages of the pumping cycle, namely the trivial, monomer, and topological stages of the control-freak protocol. Its behavior depends on the boundary conditions, the pumping protocol, and the choice of reference time. We also find that a nonzero staggered potential breaks chiral symmetry and gaps out the SSH critical point, flattening the pseudo-entropy peak that marks the phase boundary.

We further consider the Kitaev chain~\cite{Kitaev:2000nmw} to study the dependence of pseudo entropy on the choice of reference state. The Kitaev chain is a one-dimensional topological superconductor that supports Majorana boundary modes in its topological phase \cite{Alicea:2012ux,Leijnse:2012tn}. Since pseudo entropy is constructed from a transition matrix involving two states, its behavior may depend not only on the topological properties of the target state but also on the choice of reference state. In contrast to the Rice--Mele analysis, where the states are compared at different points of a pumping cycle, here we compare stationary states while keeping the superconducting target in the same even-parity sector. We consider five reference families: a parity-constrained free-fermion state, a pairing-only state, the vacuum, and two fixed gapped and entangled states. The free-fermion reference introduces filling-dependent changes, whereas the pairing-only reference becomes critical as the chain length increases. The fixed gapped references provide a control for separating reference criticality from the response near the target transitions. They also show a marked dependence on whether the half-chain subsystem has an even or odd number of sites. At $\mu=0$, we obtain an exact central result for the pairing-only reference: $S_{\mathrm{pr}}=\ln 2$ when $L=4m+2$, with an analytical finite-size correction when $L=4m$. Together with the exact vanishing of pseudo entropy for the vacuum reference, these comparisons show why a pseudo-entropy feature must be interpreted for the specified reference--target pair and subsystem partition.

The paper is organized as follows: In Sec.~\ref{sec:pseudo}, we review the concept of pseudo entropy and its formulation. In Sec.~\ref{sec:Pseudo Entropy in Thouless Pumping}, we calculate the pseudo entropy for the Thouless pump and present a detailed calculation for the Rice–Mele model using correlation matrices. We also discuss the calculation of pseudo-entropy for the Thouless pumping process and examine its behavior throughout the pumping cycle. In Sec.~\ref{sec:kietaivmodel}, we formulate pseudo entropy for the Kitaev chain using Majorana covariance matrices and correlation functions, compare five reference families with a common even-parity target, and derive the central finite-size result.

\section{Pseudo entropy}
\label{sec:pseudo}
Entanglement entropy has been used to characterize quantum phases and quantum phase transitions \cite{holzhey1994geometric,Calabrese:2004eu}. Pseudo-entropy generalizes the entanglement entropy to pairs of pure states. Given two
non-orthogonal pure states $\ket{\psi}$ and $\ket{\varphi}$ one forms
the transition matrix
\begin{equation}
  \mathcal{T}^{\psi|\varphi}
  =\frac{\ket{\psi}\bra{\varphi}}{\braket{\varphi|\psi}},
  \qquad
  \Tr\,\mathcal{T}^{\psi|\varphi}=1 ,
  \label{eq:transition}
\end{equation}
traces out a spatial complement, and evaluates the von Neumann
functional on the result \cite{ Nakata2021,Mollabashi:2021xsd,rana2026pseudoentropyquantumspin}.\\

Just as for the entanglement entropy, we split the Hilbert space as
$\mH_{\rm tot}=\mH_A\otimes\mH_B$ and obtain the reduced transition
matrix of subsystem $A$ by tracing out $B$,
\begin{equation}
  \mathcal{T}^{\psi|\varphi}_A
  \equiv\Tr_B\big[\mathcal{T}^{\psi|\varphi}\big],
  \label{eq:reduced-transition}
\end{equation}
with $\mathcal{T}^{\psi|\varphi}$ the transition matrix
\eqref{eq:transition}. The pseudo-entropy is the von Neumann
functional evaluated on this reduced matrix,
\begin{equation}
  S\big(\mathcal{T}^{\psi|\varphi}_A\big)
  =-\sum_{\lambda\in\operatorname{spec}\mathcal{T}^{\psi|\varphi}_A}
     \lambda\,\operatorname{Log}_{+}\lambda,
  \label{eq:pe-def}
\end{equation}
where the eigenvalues are counted with their multiplicities. Here $\operatorname{Log}_{+}$ denotes the endpoint logarithm with $-\pi<\arg_{+}z\le\pi$, with $\operatorname{Log}_{+}(-x)=\ln x+i\pi$ for $x>0$ and $0\operatorname{Log}_{+}0=0$. When the ordinary principal matrix logarithm exists, Eq.~\eqref{eq:pe-def} agrees with $-\Tr_A[\mathcal{T}^{\psi|\varphi}_A\Log(\mathcal{T}^{\psi|\varphi}_A)]$. On the negative real axis or at zero eigenvalues, Eq.~\eqref{eq:pe-def} instead specifies the endpoint spectral convention used in this work. When
$\ket{\psi}=\ket{\varphi}$, the transition matrix collapses to the
ordinary density matrix and \eqref{eq:pe-def} reduces to the
entanglement entropy. Away from this limit, $S$ is generally complex
because $\mathcal{T}^{\psi|\varphi}_A$ is not Hermitian. However, pseudo entropy can be real for special choices
of the initial and final states. This can occur in certain states prepared by Euclidean path
integrals~\cite{Mollabashi:2021xsd}, or when the
reduced transition matrix satisfies appropriate
reality conditions, such as those studied using
pseudo-Hermiticity~\cite{Guo:2023tjv}. It is
convenient to define the pseudo-R\'enyi entropies
\begin{equation}
  S^{(n)}\big(\mathcal{T}^{\psi|\varphi}_A\big)
  =\frac{1}{1-n}\operatorname{Log}_{+}\Tr_A\big[
     (\mathcal{T}^{\psi|\varphi}_A)^{n}\big],
  \label{eq:prenyi-def}
\end{equation}
which reduce to \eqref{eq:pe-def} in the limit
$S=\lim_{n\to1}S^{(n)}$.
With a consistent logarithm prescription in the continuation near $n=1$, the limit $S=\lim_{n\to1}S^{(n)}$ gives Eq.~\eqref{eq:pe-def}. This assumes the relevant traces are nonzero so that the pseudo-R\'enyi entropies are defined.

Exchanging the two states conjugates the transition matrix,
$\mathcal{T}^{\varphi|\psi}=(\mathcal{T}^{\psi|\varphi})^{\dagger}$. The reduced matrices obey the same relation, but their pseudo-entropies are not necessarily complex conjugates on the logarithmic branch cut. Their real parts nevertheless agree whenever the corresponding entropies are defined.
\\

We therefore define the real averaged excess pseudo-R\'enyi entropy by
\begin{equation}
\begin{aligned}
\Delta S^{(n)}
={}&\frac{1}{2}\operatorname{Re}\Big[
S^{(n)}\big(\mathcal{T}^{\psi|\varphi}_A\big)
+S^{(n)}\big(\mathcal{T}^{\varphi|\psi}_A\big)
\Big]\\
&-\frac{1}{2}\Big[
S^{(n)}\big(\rho^{\psi}_A\big)
+S^{(n)}\big(\rho^{\varphi}_A\big)
\Big].
\end{aligned}
\label{eq:real-averaged-excess-renyi}
\end{equation}
This expression is real whenever the terms are defined.
The last two terms are the ordinary R\'enyi entropies
of $\ket{\psi}$ and $\ket{\varphi}$.\\

Taking $n\to1$ gives the averaged excess entropy 
\begin{equation}
\Delta S_{12}
= \Re\,S\big(\tau_A^{1|2}\big)
-\frac{1}{2}
\left[
S\big(\rho_A^{(1)}\big)
+S\big(\rho_A^{(2)}\big)
\right].
\label{eq:dS12}
\end{equation}
Here $\ket{\Psi_1},\ket{\Psi_2}$ are the two ground states under study, $\mathcal{T}_A^{1|2}$ denotes their reduced transition matrix with state 1 in the ket, and $\rho_A^{(s)}=\Tr_B\ket{\Psi_s}\!\bra{\Psi_s}$. The real-part prescription in Eq.~\eqref{eq:dS12} does not require a complex-conjugation identity on the branch cut.
The MNT conjecture~\cite{Mollabashi:2021xsd,Mollabashi:2020yie} relates the sign of the excess pseudo entropy to the phases of the two states under particular conditions. A positive $\Delta S_{12}$ may signal a difference between the states, but its sign is not in general a reference-independent test of topology. In Sec.~\ref{sec:kietaivmodel}, positive excess also
occurs for a Majorana-trivial target compared with a
fixed gapped reference from the opposite trivial
chemical-potential regime. Both have a trivial bulk
$\mathbb{Z}_2$ invariant, although their filling
characteristics differ and the finite-size overlap
may be very small. Thus the reference state must
be specified when interpreting the sign of
$\Delta S_{12}$. \\

\section{Pseudo Entropy in Thouless Pumping}
\label{sec:Pseudo Entropy in Thouless Pumping}

In this section we study pseudo entropy in the Rice--Mele (RM) model and in the Thouless pump built from it. The RM model is the SSH chain with an added staggered on-site potential $\mu_s$ which breaks chiral symmetry and acts as a mass term, so the gap at $|v|=|w|$ stays open at $2|\mu_s|$ \cite{PhysRevLett.42.1698, Asboth2016, RevModPhys.83.1057}. Thouless pumping is the quantized transport of charge when the RM parameters $v(t)$, $w(t)$ and $\mu_s(t)$ are varied periodically and adiabatically \cite{PhysRevB.27.6083}. Because the instantaneous gap never closes, the charge pumped per cycle is an integer, given by the Chern number of the $(k,t)$ torus.
We first introduce the RM model, its Bloch Hamiltonian and band gap, and show how the pseudo-entropy between an SSH reference state and an RM state responds to chiral symmetry breaking.
Then we define the pumping cycle, first through the isolated-dimer picture and then through the full time-dependent Bloch Hamiltonian \cite{Asboth2016}.
We then compute the pseudo-entropy along both cycles under open and periodic boundary conditions and compare its behavior across the trivial, monomer, and topological stages of the pump.

\subsection{Rice–Mele model and its pseudo-entropy}\label{RM}
Consider a one-dimensional chain of $N$ unit cells with a staggered on-site potential $\mu_s$ that assigns energy $+\mu_s$ to sublattice $A$
and $-\mu_s$ to sublattice $B$. In real space, the Hamiltonian is given as \cite{Asboth2016},

\begin{align}
\hat H
&=
v\sum_{m=1}^{N}
\left(
c_{m,B}^{\dagger}c_{m,A}
+
c_{m,A}^{\dagger}c_{m,B}
\right)
\nonumber\\
&\quad
+
w\sum_{m=1}^{N-1}
\left(
c_{m+1,A}^{\dagger}c_{m,B}
+
c_{m,B}^{\dagger}c_{m+1,A}
\right)
\nonumber\\
&\quad
+
\mu_s \sum_{m=1}^{N}\left(  n_{m,A} - n_{m,B}\right), 
\label{eq:ssh_second_quantized_hamiltonian_staggared 1}
\end{align}
where $ n_{m,A} =c_{m,A}^\dagger c_{m,A} $ and $ n_{m,B} =c_{m,B}^\dagger c_{m,B}$.

To analyze the bulk properties of the model, we impose periodic boundary conditions ($c_{N+1,\alpha}\equiv c_{1,\alpha}$) and introduce the Fourier transform
\begin{equation}
c_{m,\alpha} = \frac{1}{\sqrt{N}}\sum_{k} e^{ikm}\, c_{k,\alpha}, \qquad \alpha\in\{A,B\},
\end{equation}

with $k$ running over the first Brillouin zone. Substituting into Eq.~\eqref{eq:ssh_second_quantized_hamiltonian_staggared 1}, the intracell term becomes diagonal in $k$,

\begin{equation}
v\sum_{m}\left(c_{m,B}^{\dagger}c_{m,A}+\text{h.c.}\right) = v\sum_{k}\left(c_{k,B}^{\dagger}c_{k,A}+\text{h.c.}\right),
\end{equation}

while the intercell term, which connects site $B$ in cell $m$ to site $A$ in cell $m+1$, acquires a phase factor from the unit-cell shift,

\begin{equation}
w\sum_{m}\left(c_{m+1,A}^{\dagger}c_{m,B}+\text{h.c.}\right) = w\sum_{k}\left(e^{-ik}c_{k,A}^{\dagger}c_{k,B}+\text{h.c.}\right).
\end{equation}

The staggered potential remains diagonal in both real and momentum space. Collecting terms and writing the result in the two-component spinor basis $\Psi_k^{\dagger}=(c_{k,A}^{\dagger},\,c_{k,B}^{\dagger})$, the Hamiltonian takes the block form

\begin{equation}
\hat H = \sum_k \Psi_k^{\dagger}\, H(k)\, \Psi_k, \qquad
H(k) = \begin{pmatrix} \mu_s & v+we^{-ik} \\ v+we^{ik} & -\mu_s \end{pmatrix}.
\label{eq:Hk_matrix}
\end{equation}

Writing $H(k)=h(k)\cdot\sigma$ with $\sigma=(\sigma_x,\sigma_y,\sigma_z)$ the Pauli matrices, comparison of Eq.~\eqref{eq:Hk_matrix} with $\mathbf h\cdot\sigma = \begin{pmatrix} h_z & h_x-ih_y \\ h_x+ih_y & -h_z\end{pmatrix}$ identifies the Bloch vector components as

\begin{equation}
h_x(k) = v+w\cos k, \qquad h_y(k) = w\sin k, \qquad h_z = \mu_s.
\label{eq:bloch_vector}
\end{equation}
Note that, $h_z$ is a $k$-independent constant equal to the staggered potential. Diagonalizing the $2\times2$ Hamiltonian gives the two bulk bands

\begin{align}
E_{\pm}(k) &= \pm h(k) \\
&= \pm\sqrt{h_x^2+h_y^2+h_z^2}\\
&= \pm\sqrt{v^2+w^2+2vw\cos k+\mu_s^2}.
\label{eq:dispersion}
\end{align}

When we set potential term $\mu_s=0$, we recover the familiar gapless at $|v|=|w|$ SSH dispersion $E_\pm(k)=\pm|v+we^{ik}|$. The staggered potential enters only through the constant shift $+\mu_s^2$ under the square root. The corresponding eigenstates $H(k)|u_r(k)\rangle = r\,h(k)|u_r(k)\rangle$, $r=\pm1$, are
\begin{equation}
|u_r(k)\rangle = \frac{1}{\sqrt{2h(h-rh_z)}}\begin{pmatrix} r(h_x-ih_y) \\ h-rh_z \end{pmatrix}.
\label{eq:eigenstates}
\end{equation}

The band gap is $\Delta = 2h_{\min}$, where $h_{\min}$ is the minimum of $h(k)$ over the Brillouin zone. Evaluating $h(k)$ at the two high-symmetry points,
\begin{equation}
h(0) = \sqrt{(v+w)^2+\mu_s^2}, \qquad h(\pi) = \sqrt{(v-w)^2+\mu_s^2},
\end{equation}
shows that the minimum occurs at $k_{\min}=\pi$ (for $v,w$ of the same sign), giving
\begin{equation}
h_{\min} = \sqrt{(|v|-|w|)^2+\mu_s^2}, \qquad \Delta = 2\sqrt{(|v|-|w|)^2+\mu_s^2}.
\label{eq:gap}
\end{equation}
At the SSH transition point $|v|=|w|$, the gap reduces to its minimum value $\Delta_{\min}=2|\mu_s|$, which is strictly positive whenever $\mu_s\neq0$. In this sense $\mu_s$ acts as a mass term and it gaps out what would otherwise be a gap-closing critical point, so that the transition at $|v|=|w|$ is crossed without the spectral gap ever vanishing.

When $\mu_s=0$, the Hamiltonian possesses chiral symmetry, characterized by the anticommutation relation $\mathcal{C} H \mathcal{C}^{-1} = -H$, where $\mathcal{C}=\mathrm{diag}(+1,-1,+1,-1,\ldots)$ \cite{Asboth2016}. As a consequence, the energy spectrum is symmetric about $E=0$. A finite staggered potential $\mu_s$ explicitly breaks this symmetry, causing the mid-gap edge states to acquire a finite mass and shift away from zero energy. This makes it natural to ask how sensitive the pseudo entropy is to chiral symmetry breaking. Accordingly, in this section, we examine whether pseudo entropy in the Rice-Mele model under OBC can capture the effect of chirality and what it reveals about the topological phase transition when chiral symmetry is explicitly broken.

\subsubsection{Pseudo entropy of Rice-Mele model from Correlation Matrices}
\label{sec:Fermionic_correlator}

For a generic many-body system, computing the entanglement entropy requires diagonalizing the reduced density matrix $\rho_A$, whose dimension grows exponentially with the subsystem size. For free fermions, however, the ground state is Gaussian, and Wick's theorem implies that the two-point correlation matrix determines all reduced-state properties \cite{10.21468/SciPostPhysLectNotes.54}. For a subsystem $A$, the correlation matrix is $C_{ij}=\langle c_i^\dagger c_j\rangle, \quad i,j\in A .$
If $\{\zeta_k\}$ are the eigenvalues of $C_A$, the entanglement entropy is given by the Peschel formula \cite{IngoPeschel_2003,Peschel_2009}
\begin{equation}
S_A
=
-\sum_{k=1}^{N_A}
\left[
\zeta_k\ln\zeta_k
+
(1-\zeta_k)\ln(1-\zeta_k)
\right].
\label{eq:EE}
\end{equation}

We now briefly review the corresponding construction for pseudo entropy in free-fermion systems \cite{Mollabashi:2021xsd}. Consider two Gaussian states $|\psi_1\rangle$ and $|\psi_2\rangle$, taken to be ground states of quadratic Hamiltonians
\begin{equation}
H^{(s)}
=
\sum_{i,j}h^{(s)}_{ij}c_i^\dagger c_j,
\qquad s=1,2,
\label{eq:general_H}
\end{equation}
where $c_i^\dagger$ creates a fermion in the $i$th single-particle orbital. After diagonalization, $H^{(s)}
=
\sum_k \epsilon_k^{(s)}
(\eta_k^{(s)})^\dagger \eta_k^{(s)},$
the ground state takes the Slater-determinant form $|\psi_s\rangle
=
\prod_{k=1}^{M_s}
(\eta_k^{(s)})^\dagger |0\rangle .$

Because both states are Gaussian, the transition matrix $\tau^{1|2}$ and its reduced transition matrix $\tau_A^{1|2}=\mathrm{Tr}_B\,\tau^{1|2}$ are also Gaussian. Consequently, the spectrum of $\tau_A^{1|2}$ can be extracted entirely from two-point weak-value correlators,
\begin{equation}
\wv{O}
=
\frac{\langle\psi_2|O|\psi_1\rangle}
{\langle\psi_2|\psi_1\rangle}.
\label{eq:wv}
\end{equation}
For the particle-number-conserving systems studied here, the anomalous correlators vanish, and the pseudo-correlation matrix restricted to subsystem $A$ is simply $C^{1|2}_{ij}=\wv{c_i^\dagger c_j}, \qquad i,j\in A,$
whose eigenvalues $\{\nu_k\}$ enter equation~\eqref{eq:PE_final_main}
directly. For Slater determinants, this evaluates to $C^{1|2}\big|_A = \Bigl[U_1\bigl(U_2^\dagger U_1\bigr)^{-1}U_2^\dagger\Bigr]_A,$
where $U_s$ is the $N\times M_s$ matrix of occupied single-particle orbitals of $|\psi_s\rangle$, and $[\cdots]_A$ denotes restriction to subsystem $A$.
Following Ref.~\cite{Mollabashi:2021xsd}, one constructs a pseudo-correlation matrix from these weak values. If its eigenvalues are denoted by $\{\nu_k\}$, the pseudo entropy is obtained through the fermionic correlation-matrix formula
\begin{equation}
S(\tau_A^{1|2})
=
-\sum_{k=1}^{N_A}
\left[
\nu_k\log\nu_k
+
(1-\nu_k)\log(1-\nu_k)
\right],
\label{eq:PE_final_main}
\end{equation}
where the $\nu_k$ are generally complex because $\tau_A^{1|2}$ is non-Hermitian. The corresponding pseudo-R\'enyi entropies are computed from the same spectrum. Throughout this work, the pseudo entropy is evaluated numerically using this correlation-matrix formalism.

\begin{figure}[h]
    \centering
    \includegraphics[width=1\linewidth]{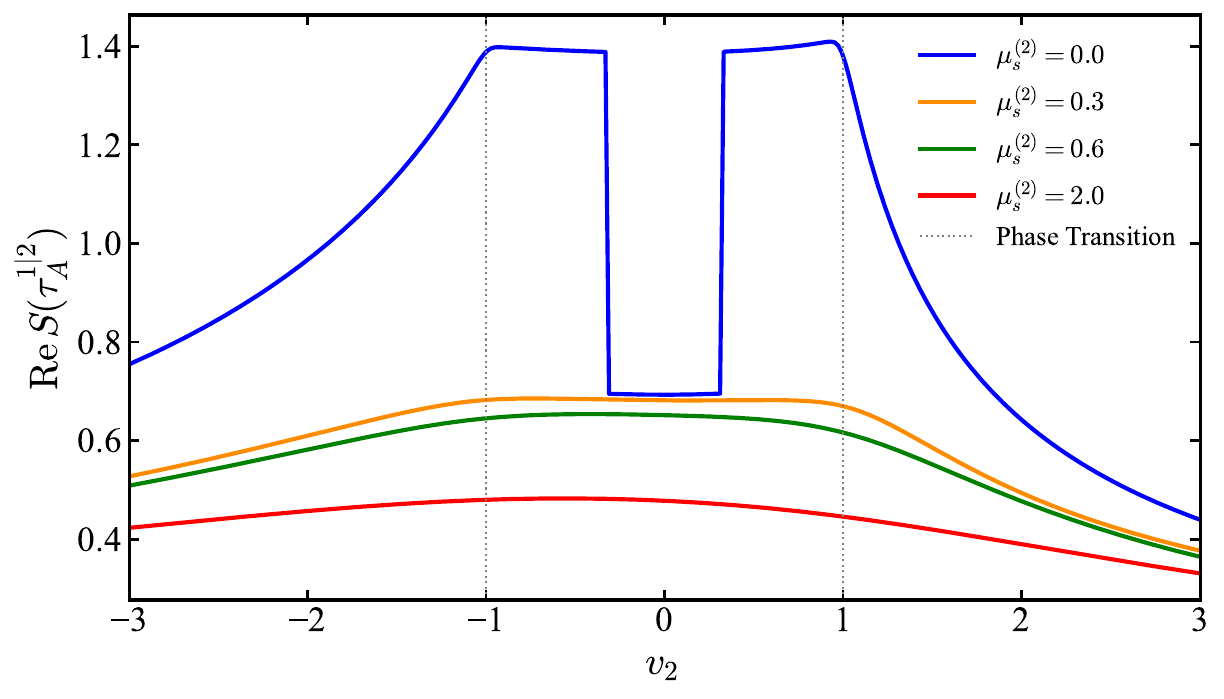}
    \caption{Real part of the pseudo entropy $\operatorname{Re} S(\tau_A^{1|2})$ as a function of the hopping $v_2$ for $N=60$ unit cells, $w=1$, with a reference state fixed in the topological phase $v_1=0.5$. The different curves correspond to staggered potentials $\mu_s = 0,\,0.3,\,0.6,\,2.0$. The vertical dashed lines at $v_2 = \pm 1$ mark the SSH phase boundaries.}
    \label{fig: rice_mele}
\end{figure}

Because the RM Hamiltonian is real symmetric for all
$\mu_s\in\mathbb{R}$, its eigenvalues are therefore real. As a result, only the real part of pseudo entropy remains, and the imaginary part is washed out. We fix state~1 in the topological SSH phase ($v_1=0.5$, $\mu_s^{(1)}=0$) and let state~2 vary across the RM parameter space with staggered potential $\mu_s^{(2)}$. In figure \ref{fig: rice_mele}, the pseudo entropy decreases monotonically with $\mu_s^{(2)}$ at every $v_2$. This is due to the gapping of the spectrum, that is, the bulk gap at the SSH critical point $v_2=w$ grows as $\Delta\simeq 2\mu_s^{(2)}$, exponentially localizing the ground-state wavefunction and reducing the subsystem correlations captured by the pseudo correlation matrix. For $\mu_s^{(2)}=0$, the pseudo-entropy exhibits a pronounced maximum near $|v_2|=w=1$, accurately tracking the SSH model's critical point. As $\mu_s^{(2)}$ increases, however, this peak becomes broader and gradually shifts away from the critical value. The staggered potential explicitly breaks chiral symmetry, thereby removing the sharp topological distinction between the phases characterized by $|v_2|<w$ and $|v_2|>w$. Consequently, the critical signature in the pseudo entropy becomes progressively smeared out. These results suggest that pseudo-entropy is sensitive to the disappearance of the topological phase boundary induced by chiral symmetry breaking. For the rest of study of Thouless pumps, we will consider $\mu_s = 0$.

\subsection{Pseudo entropy of Thouless Pumps}

By allowing the parameters of the Rice-Mele model to vary periodically and adiabatically over time, the system can operate as a quantum charge pump. This phenomenon is known as Thouless pumping \cite{Asboth2016}. The number of pumped charges is topologically protected and is mathematically governed by a Chern number. To describe this process, we change the static parameters of the RM model to time-periodic functions as the intracell hopping amplitude $v(t)$, the intercell hopping amplitude $w(t)$, and the staggered on-site potential $\mu_s(t)$. Over a driving period $T$, the time-dependent many-body Hamiltonian takes the form:

\begin{align}
\hat H(t)
&=
v(t)\sum_{m=1}^{N}
\left(
c_{m,B}^{\dagger}c_{m,A}
+
c_{m,A}^{\dagger}c_{m,B}
\right)
\nonumber\\
&\quad
+
w(t)\sum_{m=1}^{N-1}
\left(
c_{m+1,A}^{\dagger}c_{m,B}
+
c_{m,B}^{\dagger}c_{m+1,A}
\right)
\nonumber\\
&\quad
+
\mu_s(t) \sum_{m=1}^{N}\left(  n_{m,A} - n_{m,B}\right), 
\label{eq:ssh_second_quantized_hamiltonian_staggared}
\end{align}

The pump's fundamental mechanism can be intuitively understood by first isolating a single unit cell (a dimer), which corresponds to the limit in which intercell hopping is completely suppressed, $w(t)=0$. The Hamiltonian for this isolated dimer reduces to:

\begin{equation}
   \hat H_{\text{dimer}}(t) = \mu_s(t) \sigma_z + v(t) \sigma_x. 
\end{equation}

We consider a cyclic evolution of the parameters in which hopping is not allowed at the beginning ($t=0$) and at the midpoint of the cycle. If we initialize the system at $t=0$ with a strong staggered potential $\mu_s > 0$ and zero intracell hopping $v=0$, the energy offset causes the particle to heavily localize on one of the sublattices in the ground state. If we want to shift the particle from sublattice $A$ to sublattice $B$, we have to smoothly activate the intracell hopping $v(t)$ while simultaneously changing the staggered potential $\mu_s(t)$. A very standard harmonic way to write this continuous shift for a half-period is:

\begin{equation}
\mu_s(t) = \mu_0 \cos\left(\frac{\pi t}{T}\right), \quad v(t) = v_0 \sin\left(\frac{\pi t}{T}\right).
\end{equation}

During this evolution, the instantaneous energy gap is $2\sqrt{\mu_s(t)^2 + v(t)^2}$, which remains strictly open. If the parameters change slowly, a particle in the ground state will transfer its charge without jumping to the upper band.

To pump charge continuously in a 1D lattice, we need to combine the intracell shift with an intercell shift. In momentum space, applying the Fourier transform gives the time-dependent Bloch Hamiltonian,

\begin{equation}
\hat H(k,t) = \sum_k \Psi_k^{\dagger} H(k,t) \Psi_k.
\end{equation}

Writing this in the Pauli basis $H(k,t) = \mathbf{h}(k,t) \cdot \sigma$, the vector components are,

\begin{equation}
\begin{aligned}
h_x(k,t) &= v(t)+w(t)\cos k, \\
h_y(k,t) &= w(t)\sin k, \quad h_z(k,t) = \mu_s(t).
\end{aligned}
\label{eq:bloch_vector_t}
\end{equation}

During one full pumping cycle $t \in [0, T]$, the vector $\mathbf{h}(k,t)$ creates a torus in the parameter space. Because the bulk gap remains open ($\vert{}\mathbf{h}(k,t)\vert{} > 0$), exactly one integer quantum of charge is pumped per cycle.

To make this adiabatic charge pump work mathematically, we can use a piecewise linear function $f(x)$ for the interval $x \in [0, 1)$ as described in \cite{Asboth2016} as,

\begin{equation}
f(x) = \begin{cases}
8x, & \text{if } x \le 1/8 \\
1, & \text{if } 1/8 \le x < 3/8 \\
1 - 8(x - 3/8), & \text{if } 3/8 \le x < 1/2 \\
0, & \text{otherwise}
\end{cases}
\label{eq: control freak}
\end{equation}

By adding phase shifts to this function, the time-dependent parameters for one full period ($0 \le t < T$) are defined as:

\begin{equation}
\begin{aligned}
\mu_s(t) &= f(t/T) - f(t/T + 1/2), \quad v(t) = 2f(t/T + 1/4), \\
w(t) &= f(t/T - 1/4).
\end{aligned}
\label{eq:cf_protocol}
\end{equation}

Because of this specific timing, the system passes through different phases. At integer times ($t/T = n$), it is in the trivial SSH phase. At quarter-periods ($t/T = n + 1/4$), hopping becomes zero, creating disconnected monomers. Finally, at half-periods ($t/T = n + 1/2$), the system goes into the non-trivial topological SSH phase.

Throughout the entire cycle, the Hamiltonian in momentum space is:

\begin{equation}
\hat H(k,t) = (v(t) + w(t)\cos k)\sigma_x + w(t)\sin k \sigma_y + \mu_s(t)\sigma_z.
\end{equation}

Since the energy gap never closes, the system smoothly moves a particle from site $\vert{}m, A\rangle$ to $\vert{}m, B\rangle$, and then to $\vert{}m+1, A\rangle$ adiabatically, creating a perfect quantized charge pump.

\begin{figure*}[tbp]
    \centering

    \begin{subfigure}{\columnwidth}
        \centering
         \includegraphics[width=1\linewidth]{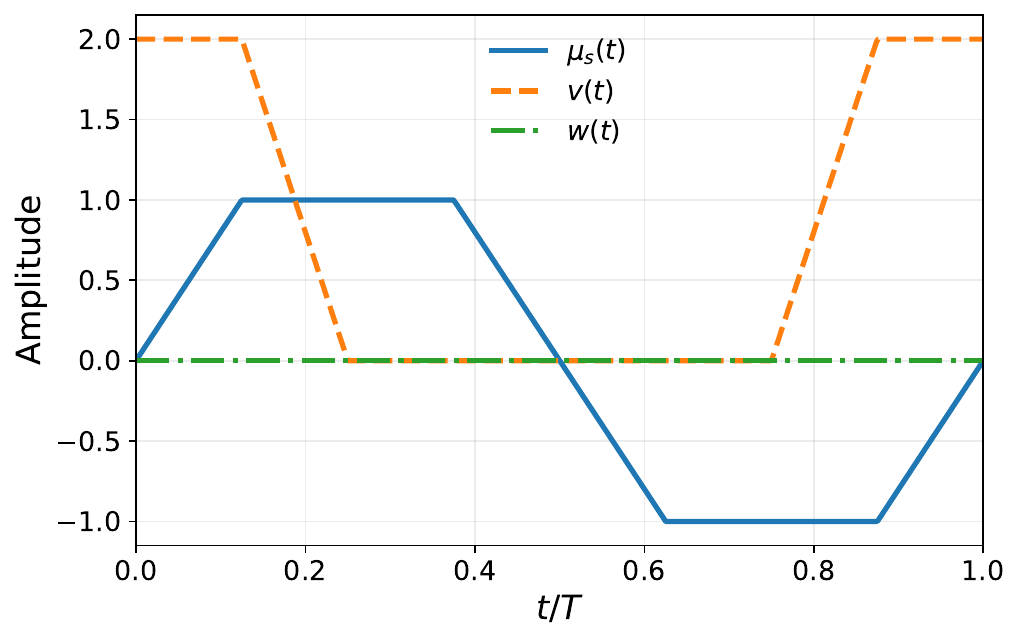}
        \caption{Control freak pumping}
        \label{fig: CF}
    \end{subfigure}
    \hfill
    \begin{subfigure}{\columnwidth}
        \centering
          \includegraphics[width=1\linewidth]{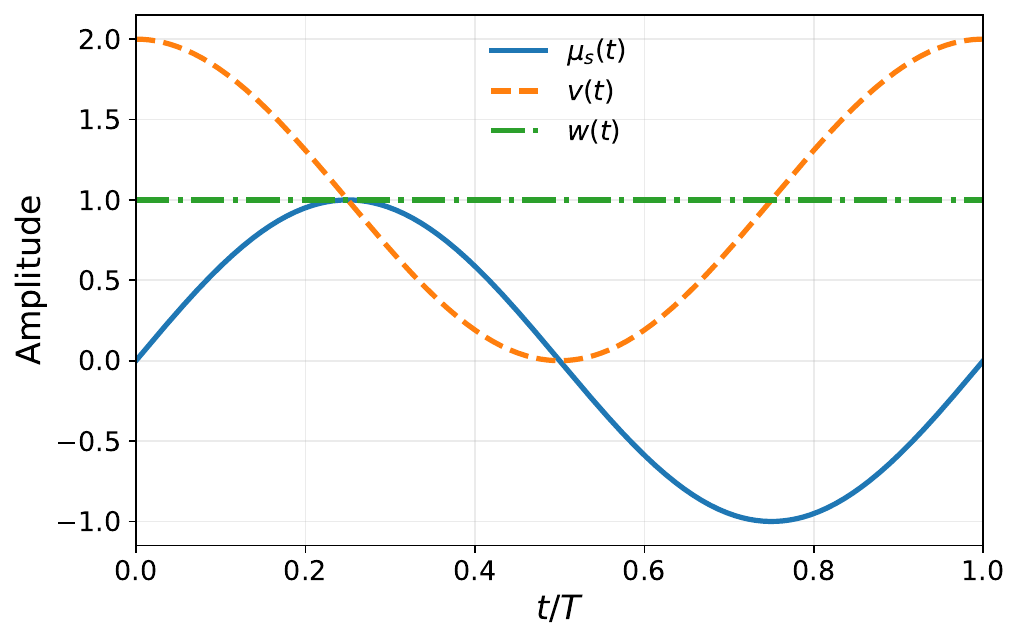}
        \caption{Smooth Pumping}
        \label{fig:SM}
    \end{subfigure}

   \caption{Driving parameters $\mu_s(t)$, $v(t)$ and $w(t)$ over one pumping cycle for the (a) control-freak and (b) smooth protocols, the latter with $\bar v=1$. Both cycles start in the trivial SSH phase and reach the non-trivial phase at $t/T=0.5$, while $\mu_s(t)\neq0$ whenever $v(t)=w(t)$, keeping the gap open.}
    \label{fig:CF SM}
\end{figure*}

The control-freak protocol is convenient because the particle hops between well-separated sites, but it is not essential. For a generic periodically driven insulator, the eigenstates are delocalized over the bulk, so the simple picture of a particle moving from site to site no longer applies. Hence, it is necessary to argue through adiabatic deformations instead. We again take the RM model, now with a smooth modulation sequence with angular frequency $\Omega=2\pi/T$ \cite{Asboth2016},

\begin{equation}
\mu_s(t) = \sin\Omega t, \qquad v(t) = \bar v + \cos\Omega t, \qquad w(t) = 1,
\label{eq:smooth}
\end{equation}

where the average intracell hopping $\bar v$ is the only free parameter of the sequence. In the $(h_x,h_y,h_z)$ space, the Bloch vector $\mathbf h(k,t)$ of the smooth pump again traces out a torus over one cycle in $k$ and $t$. The time evolution of driving parameters from equations \ref{eq:cf_protocol} and \ref{eq:smooth} is shown in figures \ref{fig: CF} and \ref{fig:SM} respectively.

We now compute the pseudo entropy along the pumping cycle using the averaged excess entropy $\Delta S_{12}=\mathrm{Re}\,S(\tau_A^{1|2})-\tfrac12\left(S(\rho_1)_A+S(\rho_2)_A\right)$. At each instant the half-filled ground state of $\hat H(t)$ is obtained by diagonalizing the single-particle Hamiltonian and filling the lowest $N$ orbitals, where $N$ is the number of unit cells. The subsystem $A$ is the left half of the chain. Here the reference state is $|\psi(t_{\mathrm{ref}})\rangle$, fixed at one point of the cycle, while the target state $|\psi(t)\rangle$ is swept over $t/T\in[0,1]$. We choose reference times in the non-trivial SSH phase ($t_{\mathrm{ref}}/T=0.45,\,0.48,\,0.52,\,0.55$) and in the trivial SSH phase ($t_{\mathrm{ref}}/T=0.00,\,0.05,\,0.10,\,0.90,\,0.95$) throughout this section.

\begin{figure*}[tbp]
    \centering

    \begin{subfigure}{\columnwidth}
        \centering
       \includegraphics[width=1\linewidth]{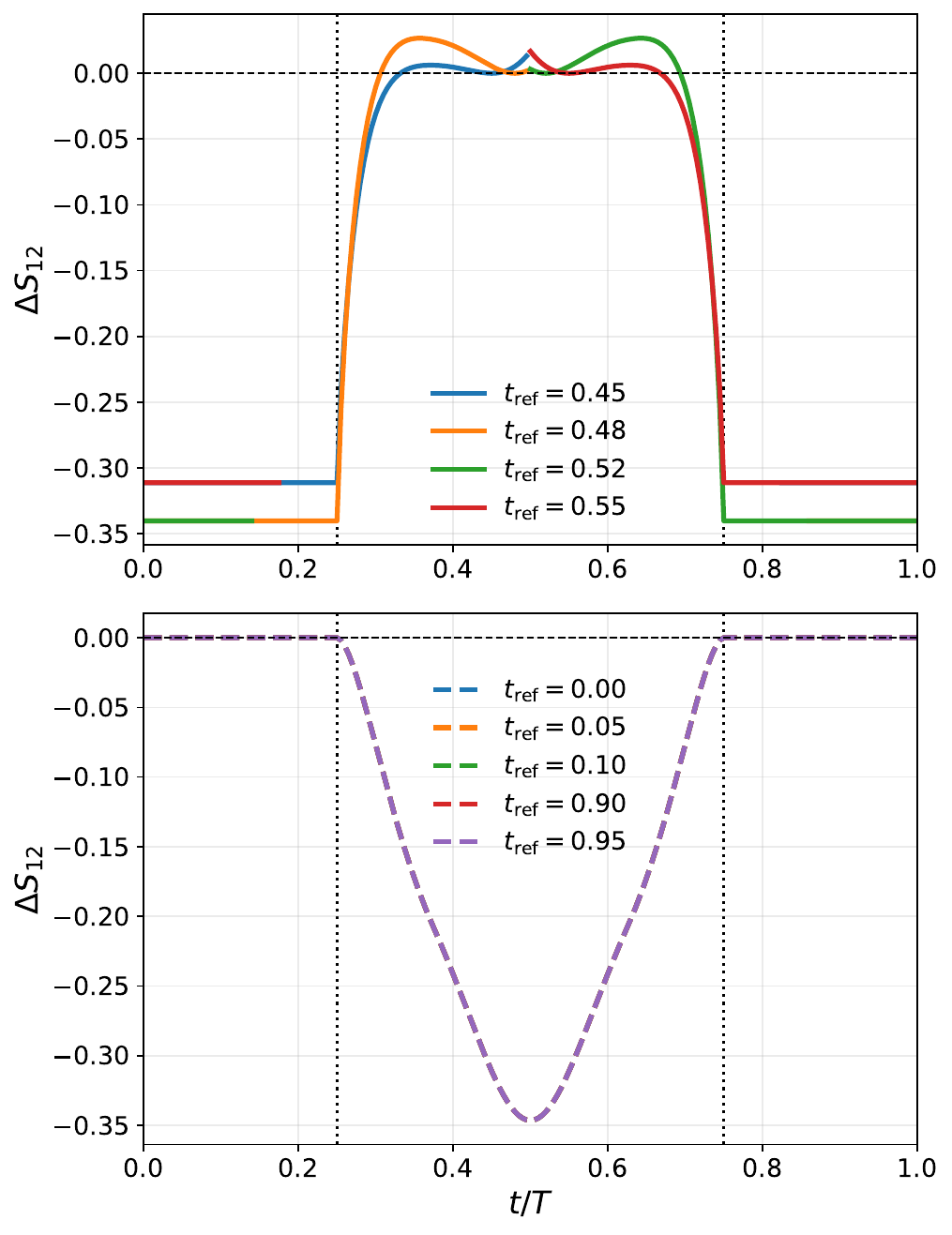}
        \caption{Control-freak pump under open boundary conditions.}
        \label{fig:OBC CF}
    \end{subfigure}
    \hfill
    \begin{subfigure}{\columnwidth}
        \centering
        \includegraphics[width=1\linewidth]{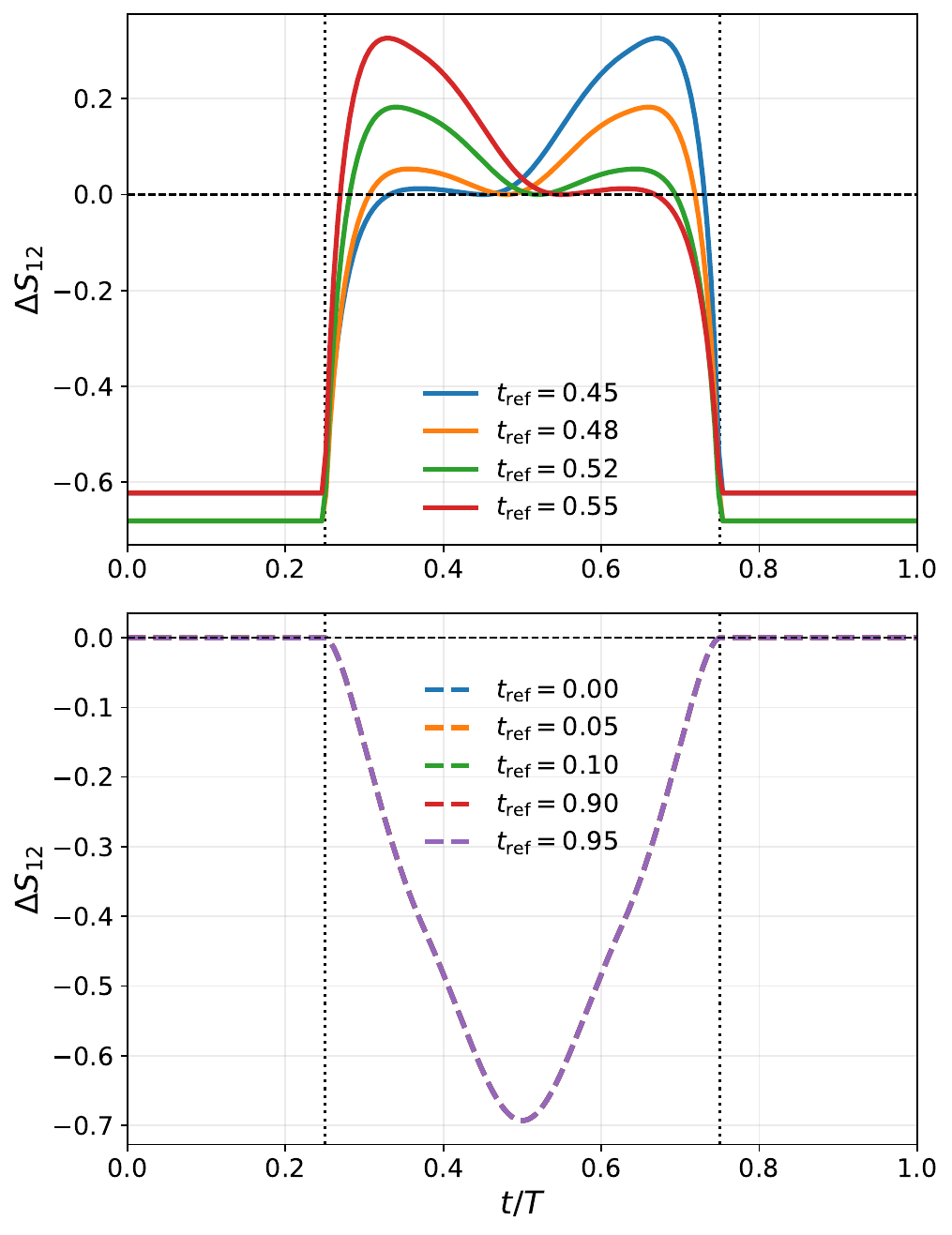}
        \caption{Control-freak pump under periodic boundary conditions.}
        \label{fig:PBC CF}
    \end{subfigure}

   \caption{$\Delta S_{12}$ as a function of $t/T$ for the control-freak Rice--Mele pump, with $N=60$ unit cells. In each panel, the upper plot uses reference times in the non-trivial SSH phase and the lower plot uses reference times in the trivial SSH phase. Points where the overlap with the reference vanishes are omitted. (a) For non-trivial references, $\Delta S_{12}$ sits on a negative plateau outside $t/T\in[0.25,0.75]$ and rises to near zero inside this window, with small positive peaks. For trivial references, all curves collapse onto a single curve that vanishes outside this window and forms a smooth negative dip at $t/T=0.5$. (b) The curves have the same pattern but with roughly double the magnitude.}
    \label{fig:pes_saturation_gapped}
\end{figure*}

For control freak limit the parameters $\mu_s(t)$, $v(t)$ and $w(t)$ follow the piecewise-linear protocol built from $f(x)$ of Eq.~\ref{eq: control freak}. The cycle passes through the trivial phase ($t/T=0,1$), the disconnected-monomer points ($t/T=1/4,\,3/4$) and the non-trivial phase ($t/T=1/2$), so $\Delta S_{12}$ tracks how the instantaneous state moves between these regimes relative to the chosen reference. Under OBC Fig.~\ref{fig:OBC CF}, $\Delta S_{12}$ for non-trivial references stays on a negative plateau outside $t/T\in[0.25,0.75]$ and turns positive inside this window, while for trivial references all curves collapse onto one curve that vanishes outside the window and dips negative at $t/T=0.5$. Under PBC Fig.~\ref{fig:PBC CF}, the pattern is the same but the magnitudes are roughly twice as large. The calculation was done for $N=60$ unit cells.

\begin{figure*}[tbp]
    \centering

    \begin{subfigure}{\columnwidth}
        \centering
       \includegraphics[width=1\linewidth]{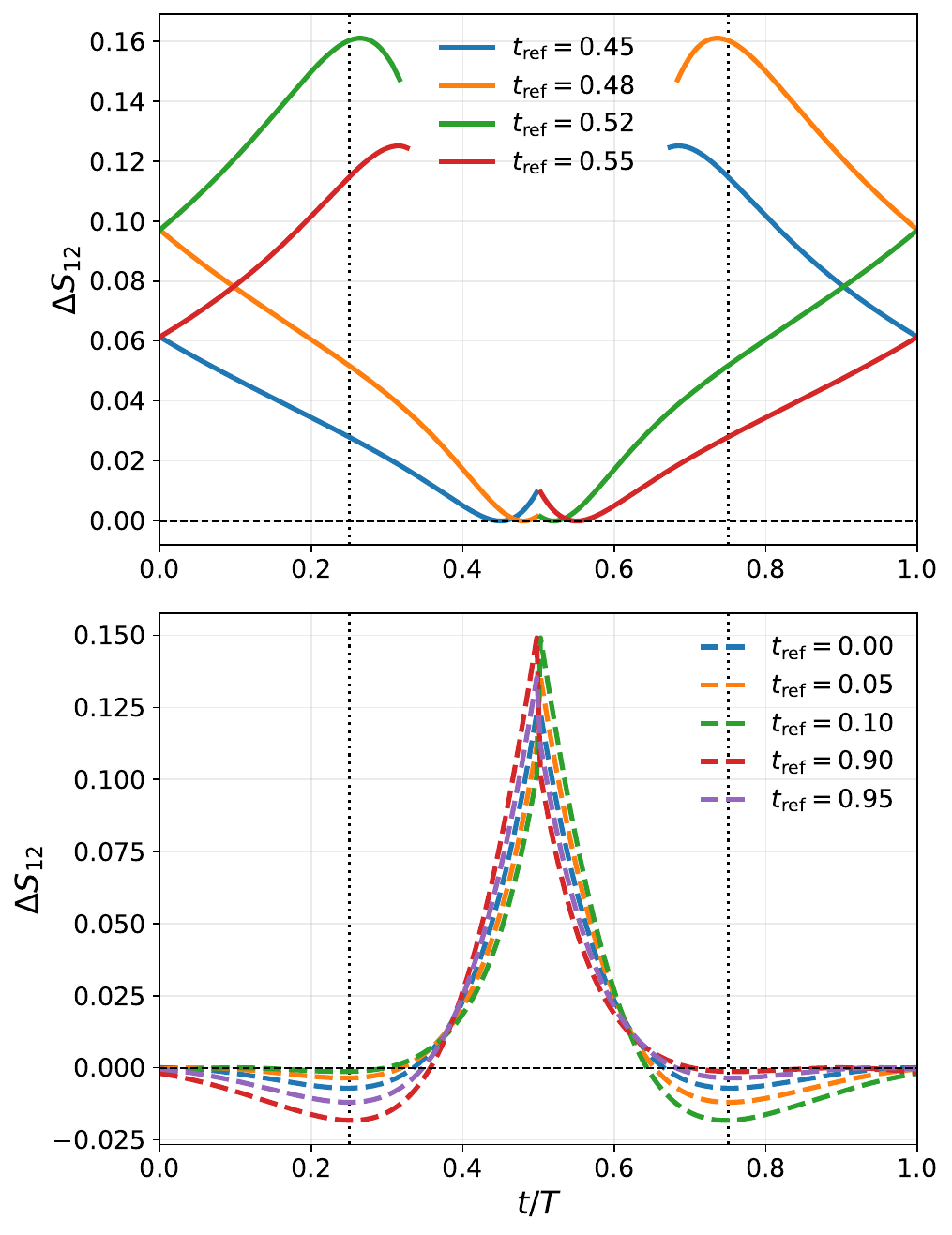}
        \caption{OBC Smooth Pump}
        \label{fig:OBC SM}
    \end{subfigure}
    \hfill
    \begin{subfigure}{\columnwidth}
        \centering
        \includegraphics[width=1\linewidth]{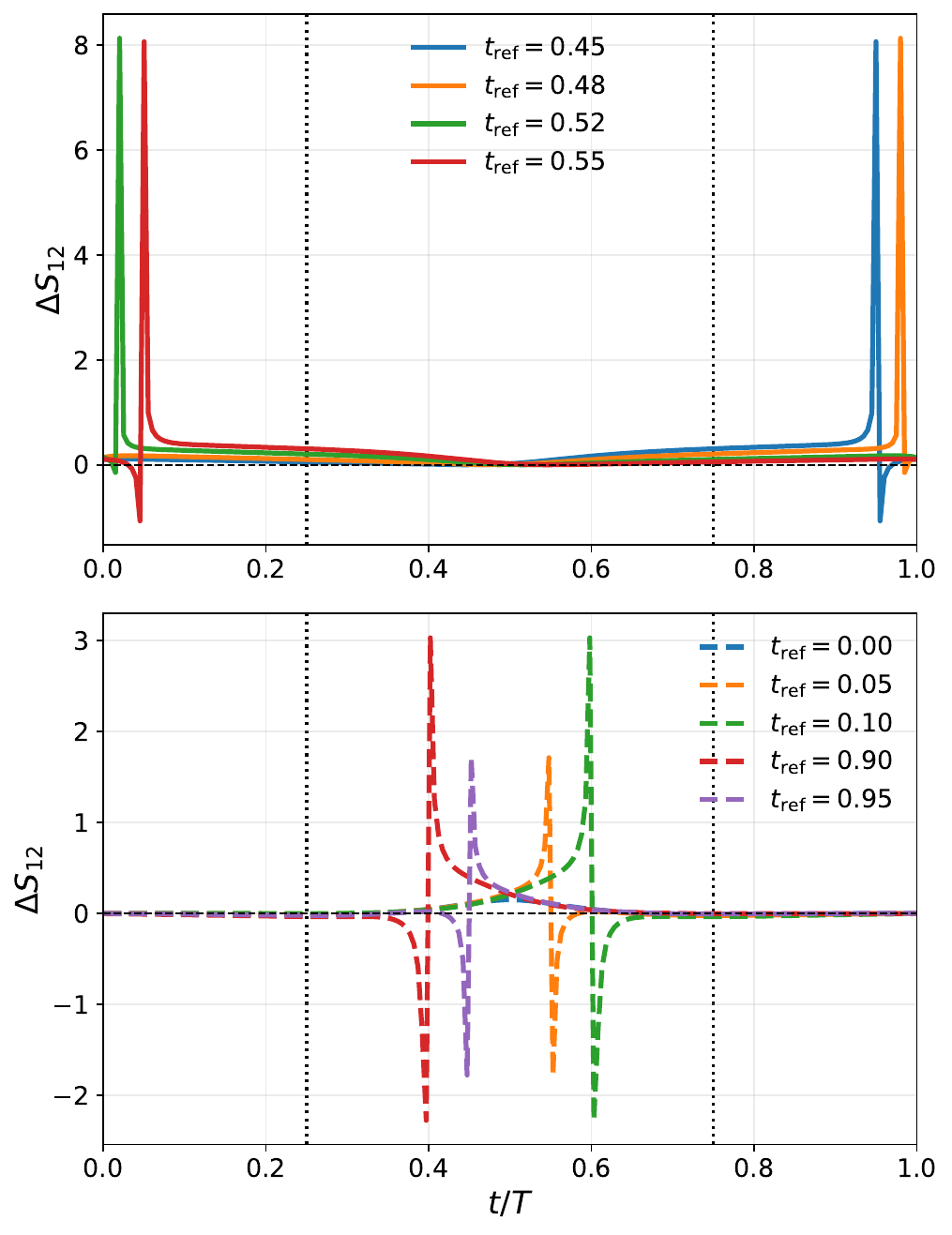}
        \caption{PBC Smooth Pump}
        \label{fig:PBC SM}
    \end{subfigure}

   \caption{Pseudo entropy difference $\Delta S_{12}$ versus $t/T$ for the smooth Rice--Mele pump with $N=60$ unit cells. Upper panels use reference times in the non-trivial SSH phase, lower panels in the trivial SSH phase. (a) For non-trivial references $\Delta S_{12}$ is positive and vanishes at $t/T=0.5$, with gaps where the overlap is singular. For trivial references the curves coincide, with a sharp peak at $t/T=0.5$ and shallow negative dips near $t/T=0.25$ and $0.75$. (b) $\Delta S_{12}$ is small and smooth except for sharp spikes at $t/T\simeq t_{\rm ref}\pm0.5$, which are divergences at zero overlap.}
    \label{fig:smooth pumping pseudo entropy}
\end{figure*}

Similarly, for smooth pump, the parameters follow the harmonic protocol as described in Eq.~\ref{eq:smooth}. The cycle therefore starts and ends in the trivial phase ($v=2,\,w=1$) and reaches the non-trivial phase at $t/T=1/2$ ($v=0,\,w=1$). The instantaneous state changes continuously and the pseudo entropy has no points where the Hamiltonian decouples since the parameter never switch off unlike control freak limit. We use exact same reference times and definition of $\Delta S_{12}$ as above, with $N=60$ unit cells. Under OBC Fig.~\ref{fig:OBC SM}, $\Delta S_{12}$ is positive for non-trivial references and goes to zero at $t/T=0.5$, whereas for trivial references it stays close to zero except for a sharp peak at $t/T=0.5$. Under PBC (Fig.~\ref{fig:PBC SM}), $\Delta S_{12}$ is small and smooth over most of the cycle, apart from sharp spikes near $t/T=t_{\rm ref}\pm0.5$. These are not features of the entropy itself. At $k=\pi$ the Bloch vector becomes antipodal to that of the reference, so the overlap vanishes exactly and $\Delta S_{12}$ diverges. For $\bar v=1$ this happens at $t=t_{\rm ref}\pm T/2$. The height of a spike depends on the time grid, and $\Delta S_{12}$ changes sign across it for both trivial and non-trivial references. 

Taken together, $\Delta S_{12}$ distinguishes the trivial, monomer and non-trivial stages of the control-freak Rice--Mele cycle, and the trivial and non-trivial stages of the smooth cycle, which has no monomer point. In both protocols its shape depends on the boundary condition, the protocol and the reference time. This suggests that excess pseudo entropy can serve as a diagnostic of the pumping cycle, provided the reference state is specified.

\section{Pseudo entropy in Majorana phases}
\label{sec:kietaivmodel}

In this section, we examine pseudo entropy in a one-dimensional Kitaev chain topological superconductor\cite{Kitaev:2000nmw,Alicea:2012ux,Leijnse:2012tn}. The Kitaev model  provides a natural setting for studying the reference-state dependence of pseudo entropy. In the topological phase\cite{Kitaev:2000nmw,Lutchyn:2010xey,Oreg:2010xfn}, an open chain supports Majorana boundary modes, and the corresponding ground states can be conveniently analyzed within the Gaussian-state formalism. Unlike the pumping analysis of the previous section, the comparative analysis is between the stationary states. 

The Kitaev chain is a one-dimensional tight-binding
model of spinless fermions with $p$-wave superconducting pairing\cite{Kitaev:2000nmw,Read:1999fn}. The Kitaev chain is modeled by a Hamiltonian with $L$ spinless fermionic sites as \cite{Kitaev:2000nmw}:
\begin{align}
 H(\mu)=&-t\sum_j\left(c_j^\dagger c_{j+1}+c_{j+1}^\dagger c_j\right)
 -\mu\sum_j\left(c_j^\dagger c_j-\frac12\right)
 \nonumber\\
 &+\Delta\sum_j\left(c_jc_{j+1}+c_{j+1}^\dagger c_j^\dagger\right),
 \label{eq:kitaev}
\end{align}
where $t$ and $\Delta$ are real and nonzero. The first term corresponds to a hopping term where $t$ represents the hopping strength rather than
the pump time of
Sec.~\ref{sec:Pseudo Entropy in Thouless Pumping}, the second term is the chemical potential, while the third term corresponds to superconducting pairing. For open boundary conditions (OBC), the chain ends at site $L$; for periodic or anti-periodic boundary conditions (PBC/APBC), we connect the boundary sites as $c_{L+1}= \pm c_1$, respectively. We use periodic or anti-periodic boundary conditions to describe the bulk phase structure, while the reference-state analysis is performed with open boundary conditions to retain the Majorana edge modes. Through a Jordan--Wigner transformation, the Kitaev chain maps onto the
anisotropic XY spin chain in a transverse field, which reduces to the
transverse-field Ising chain at $t=\Delta$~\cite{Lieb:1961fr}

 Here and throughout, we introduce the index
$s=1,2$ that label the two parent Kitaev-chain Hamiltonians compared in the pseudo-entropy transition operator. Because of the superconducting pairing term, we introduce the Nambu spinor
$\Psi_k=(c_k,c_{-k}^\dagger)^T$. The corresponding paired momentum block is
\begin{equation}
\mathcal{H}_s(k)
=\xi_s(k)\tau_z+2\Delta\sin k\,\tau_y,
\qquad
\xi_s(k)=-\mu^{(s)}-2t\cos k.
\end{equation}
with spectrum $\pm E_s(k)$,
\begin{equation}
 E_s(k)=\sqrt{\xi_s(k)^2+4\Delta^2\sin^2 k}.
 \label{eq:kit-bulk-dispersion}
\end{equation}

The quadratic Hamiltonian can then be written in Bogoliubov--de Gennes form as
\begin{equation}
H_s=\frac{1}{2}\sum_k
\Psi_k^\dagger\mathcal{H}_s(k)\Psi_k
+\mathrm{const.}
\end{equation}
The factor $1/2$ compensates for the double counting
of the paired momenta $k$ and $-k$ in the Nambu representation.

For $t>0$, the bulk gap closes when $E_s(k)=0$, which requires both
$\sin k=0$ and $\xi_s(k)=0$. This occurs at
$(\mu,k)=(-2t,0)$ and $(2t,\pi)$. These critical points separate the
topological phase $|\mu|<2t$ from the trivial phases $|\mu|>2t$; in the
topological regime, $\xi(0)$ and $\xi(\pi)$ have opposite signs, corresponding
to a nontrivial winding of the BdG vector \cite{Kitaev:2000nmw}. Under OBC, the topological phase supports Majorana modes localized at the two ends of the chain\cite{Kitaev:2000nmw,Beenakker:2011np,Aguado:2017}.

It is useful to parametrize the BdG Hamiltonian by the angle
$\theta_s(k)$ of its normalized Bloch vector in the $y$--$z$ plane,
\begin{equation}
 \cos\theta_s(k)=\frac{\xi_s(k)}{E_s(k)},
 \qquad
 \sin\theta_s(k)=\frac{2\Delta\sin k}{E_s(k)}.
 \label{eq:theta}
\end{equation} 
For $0<k<\pi$, an even-parity BCS ground-state can be written in the standard Bogoliubov form \cite{Kitaev:2000nmw,Bernevig:2013,Ring:1980}
\begin{equation}
 \begin{aligned}
 \ket{\Omega_s(k)}&=
 u_{s,k}\ket{00}_{k,-k}+v_{s,k}\ket{11}_{k,-k},\\
 u_{s,k}&=\cos\frac{\theta_s(k)}2,
 \qquad
 v_{s,k}=\mathrm{i}\sin\frac{\theta_s(k)}2.
 \end{aligned}
 \label{eq:bcs}
\end{equation}
The overlap of two such blocks is
\begin{equation}
 d_k\equiv\braket{\Omega_2(k)|\Omega_1(k)}
 =\cos\frac{\theta_1(k)-\theta_2(k)}{2}.
 \label{eq:modeoverlap}
\end{equation}
For paired momentum sectors, the many-body overlap factorizes into contributions of the form $d_k$. Therefore, a zero of $d_k$ makes the pseudo-entropy construction ill-defined. This issue will be addressed in the open-chain parity analysis below.

\subsection{Gaussian formulation of pseudo entropy}

The Kitaev chain is quadratic in fermionic operators, so its Gaussian
states can be characterized by two-point correlations. Unlike the
number-conserving Rice--Mele model, superconducting pairing requires both normal and anomalous contractions. For the computation of entanglement and pseudo-entropy, we use Majorana operators
\begin{equation}
 \gamma_{2j-1}=c_j+c_j^\dagger,\qquad
 \gamma_{2j}=-i(c_j-c_j^\dagger),
 \label{eq:iv-majoranas}
\end{equation}
and consider two pure Gaussian states $|\Omega_1\rangle$ and
$|\Omega_2\rangle$ with nonzero overlap
$d=\langle\Omega_2|\Omega_1\rangle$. Here State 1 denotes the reference
and state 2 the target. We define the normalized Majorana transition covariance as ~\cite{10.21468/SciPostPhysLectNotes.54,Peschel_2009,Vidal:2002rm,Latorre:2003kg,Bravyi:2004bud}
\begin{equation}
\Gamma^{1|2}_{mn}=\frac{\mathrm{i}}{2}
\frac{\langle\Omega_2|[\gamma_m,\gamma_n]|\Omega_1\rangle}{d}.
\label{eq:kit-transition-covariance}
\end{equation}

For a subsystem $A$ of $\ell$ sites, let $\Gamma_A^{1|2}$ be the
corresponding $2\ell\times2\ell$ principal submatrix and define
$M_A^{1|2}=\mathrm{i}\Gamma_A^{1|2}$. As in the ordinary Gaussian case, its
spectrum occurs in pairs,
\begin{equation}
 \operatorname{spec}M_A^{1|2}
 =
 \ \{\pm\nu_1,\ldots,\pm\nu_\ell\}
 \label{eq:pairedspectrum}
\end{equation}

The transition eigenvalues can be complex and are not restricted to the interval $[-1,1]$. With $p_{\alpha,\pm}=(1\pm\nu_\alpha)/2$,
the additive Gaussian expression is
\begin{equation}
 S_{\mathrm{mode}}
 =-\sum_{\alpha=1}^{\ell}\sum_{\eta=\pm}
 p_{\alpha,\eta}\operatorname{Log}_{+}p_{\alpha,\eta}.
 \label{eq:iv-mode-entropy}
\end{equation}

Here $\operatorname{Log}_{+}$ denotes the logarithm with $-\pi<\arg_{+}z\leq\pi$, where negative real arguments are assigned the upper side of the branch cut. For identical states,
Eq.~\eqref{eq:iv-mode-entropy} reduces to the ordinary entanglement entropy in Eq.~\eqref{eq:EE}. In the Gaussian normal form, let $p_\alpha=p_{\alpha,+}$ then the many-body transition
eigenvalues are products of the mode weights,
\begin{equation}
 q_{\boldsymbol n}
 =\prod_{\alpha=1}^{\ell}
 p_\alpha^{n_\alpha}(1-p_\alpha)^{1-n_\alpha},
 \qquad n_\alpha\in\{0,1\}.
 \label{eq:refdep-products}
\end{equation}
The endpoint principal-spectrum pseudo entropy is then
\begin{equation}
S_{\mathrm{pr}}(\mathcal{T}_A^{1|2})
=-\sum_{\boldsymbol n}
q_{\boldsymbol n}
\operatorname{Log}_{+}q_{\boldsymbol n}.
\label{eq:refdep-principal}
\end{equation}
where $-\pi<\operatorname{Im}\operatorname{Log}_{+}z\leq\pi$ and $0\operatorname{Log}_{+}0=0$. The assignment
$\operatorname{Log}_{+}(-x)=\ln x+\mathrm{i}\pi$ for $x>0$ defines an endpoint spectral extension on the negative real axis.
 \\

For distinct states, the principal logarithm of a product is not generally equal to the sum of the principal logarithms of its factors. Therefore, the additive mode expression in Eq.~\eqref{eq:iv-mode-entropy} and the principal-spectrum definition in Eq.~\eqref{eq:refdep-principal} need not give the same complex pseudo entropy. Throughout this work, we use Eq.~\ref{eq:refdep-principal} as the endpoint prescription.
This prescription is applied independently to each reference--target comparison and should not be interpreted as a continuous logarithmic continuation along the parameter scan. For real mode weights, the real parts of Eqs.~\eqref{eq:iv-mode-entropy} and \eqref{eq:refdep-principal} agree, but their imaginary parts can differ. For complex mode weights, both parts can differ.\\

For real $p_\alpha$, both parts of Eq.~\eqref{eq:refdep-principal}
can be evaluated from the $\ell$ independent modes, without
constructing the $2^\ell$ many-body products:
\begin{equation}
\begin{aligned}
\operatorname{Re}S_{\mathrm{pr}}
&=-\sum_{\alpha=1}^{\ell}
\left[
p_\alpha\ln|p_\alpha|
+(1-p_\alpha)\ln|1-p_\alpha|
\right],\\
\operatorname{Im}S_{\mathrm{pr}}
&=\frac{\pi}{2}
\left[
\prod_{\alpha=1}^{\ell}
\bigl(|p_\alpha|+|1-p_\alpha|\bigr)-1
\right].
\end{aligned}
\end{equation}
Here $x\ln|x|$ is set to zero at $x=0$ by continuity.
The second identity uses the upper-bank convention for negative
many-body weights and follows from
$\sum_{\boldsymbol n}q_{\boldsymbol n}=1$ and
$\sum_{\boldsymbol n}|q_{\boldsymbol n}|=
\prod_\alpha(|p_\alpha|+|1-p_\alpha|)$.
Numerically the product can be evaluated using a sum of logarithms.
These formulas require real independent mode weights; complex
weights must instead be treated with the full endpoint prescription.

To evaluate the transition covariance for the open chain, we
introduce the real-space Nambu spinor
\begin{equation}
\Psi =
\left(
c_1,\ldots,c_L,
c_1^\dagger,\ldots,c_L^\dagger
\right)^T .
\label{eq:nambu_realspace}
\end{equation}

The BdG matrix for the open chain  Hamiltonian introduced in Eq.~\eqref{eq:kitaev} is
\begin{equation}
 H_{\mathrm{BdG}}=
 \begin{pmatrix}h&B\\-B&-h\end{pmatrix},
 \qquad h^T=h,\quad B^T=-B,
 \label{eq:refdep-bdg}
\end{equation}
where $h_{jj}=-\mu$, $h_{j,j+1}=h_{j+1,j}=-t$, and
$B_{j,j+1}=-\Delta$. For a pure Gaussian state $|\Omega_s\rangle$, we collect its occupied Nambu vectors in a $2L\times L$ frame $U_s$ with $U_s^\dagger U_s=I_L$ and $U_s^T\tau_x U_s=0$, where
$\tau_x=\left(\begin{smallmatrix}0&I_L\\I_L&0\end{smallmatrix}\right)$.
For nondegenerate modes, the negative-energy eigenvectors determine the absolute ground-state frame. The required parity sector and any unresolved zero mode are selected separately, as discussed in the next subsection.\\

Using the weak-value notation introduced in Eq.~\ref{eq:wv}, the generalized
Wick theorem gives the transition correlator
\cite{Balian:1969tb,Ring:1980}
\begin{align}
 G^{1|2}_{ij}&={}_2\langle
 \boldsymbol\Psi_j^\dagger\boldsymbol\Psi_i\rangle_1,\nonumber\\
 G^{1|2}&=U_1(U_2^\dagger U_1)^{-1}U_2^\dagger
 \label{eq:refdep-G}
\end{align}

Since particle number is not conserved in the Kitaev chain, $G^{1|2}$ contains both normal and anomalous contractions. In particular, the weak values of $c_j^\dagger c_i$, $c_j c_i$, and $c_j^\dagger c_i^\dagger$ must all be retained. Unlike an ordinary density matrix, the two anomalous transition contractions are not related by Hermitian conjugation. \\

For a subsystem $A$ containing $\ell$ sites, we retain both Nambu
components of each site and define
\begin{equation}
 G_A^{1|2}=[G^{1|2}]_A .
\end{equation}
The canonical anticommutation relations imply
\begin{equation}
 \tau_{x,A}(G_A^{1|2})^T\tau_{x,A}
 =I_{2\ell}-G_A^{1|2},
 \qquad
 \tau_{x,A}=
 \begin{pmatrix}
 0&I_\ell\\
 I_\ell&0
 \end{pmatrix}.
\end{equation}
Its eigenvalues therefore occur in complementary pairs,
\begin{equation}
 \operatorname{spec}G_A^{1|2}
 =\{p_1,1-p_1,\ldots,p_\ell,1-p_\ell\}.
 \label{eq:refdep-paired}
\end{equation}
 \\
The $2\ell$ Nambu eigenvalues therefore describe only
$\ell$ independent fermionic modes, with each pair
$p_\alpha$ and $1-p_\alpha$ corresponding to the same
effective mode. The transpose in the particle--hole relation is essential, since the transition correlator is generally non-Hermitian. The Nambu doubling is therefore removed when counting the independent mode contributions.\\

To verify the Gaussian formulation, we compared the results with direct many-body calculations for \(L=4,6,8\), using the same fermion-parity selection in both approaches. In 224 valid comparisons, the real and imaginary parts of the pseudo entropy agreed within numerical precision, with maximum differences of \(3.3\times10^{-13}\) and \(4.1\times10^{-15}\), respectively. The transition correlators and reduced transition operators were also found to be consistent.

\subsection{Reference-state dependence}
\label{sec:families}

The Rice--Mele pumping cycle illustrates that pseudo entropy depends on both states of the transition operator. We now examine this dependence in the Kitaev chain by changing the structure of the reference Hamiltonian while retaining
the same even-parity superconducting target state at each $\mu$. This provides a complementary
test of whether the observed pseudo-entropy features are intrinsic to
the target transition or depend on the reference state as well. 

For this analysis, we use an open chain of even length $L$, a subsystem
$A=\{1,\ldots,\ell\}$ with $\ell=L/2$, and the Hamiltonian
\begin{align}
 H(t,\mu,\Delta)={}&-t\sum_{j=1}^{L-1}
 (c_j^\dagger c_{j+1}+c_{j+1}^\dagger c_j)
 -\mu\sum_{j=1}^{L}(n_j-\tfrac12)
 \nonumber\\
 &+\Delta\sum_{j=1}^{L-1}
 (c_jc_{j+1}+c_{j+1}^\dagger c_j^\dagger).
 \label{eq:refdep-H}
\end{align}

Throughout the comparison, we choose the same target state,
\begin{equation}
|\Psi_+(\mu)\rangle
=\operatorname{GS}_{P=+1}[H(1,\mu,1)],\qquad
P=(-1)^{\sum_{j=1}^{L}n_j}.
\label{eq:refdep-target}
\end{equation}
Here the target has $t=\Delta=1$, with bulk transitions at
$\mu=\pm2t$. The even-parity condition fixes the target
independently of the reference state. For even $L$, this is the
absolute ground state at $\mu\ne0$. At $\mu=0$, it selects the
even state within the Majorana ground-state doublet.\\

We compare five reference families. The first three are:
\begin{align}
 |\Psi_t(\mu)\rangle&=\operatorname{GS}_{P=+1}[H(1,\mu,0)],
       &&\text{free-fermion reference},\nonumber\\
 |\Psi_\Delta\rangle&=\operatorname{GS}[H(0,0,1)],
       &&\text{pairing-only reference},\nonumber\\
 |0\rangle&=\bigotimes_{j=1}^{L}|0_j\rangle,
       &&\text{vacuum reference}.
 \label{eq:refdep-references}
\end{align}

The remaining two are fixed, gapped reference states, introduced below.\\

These choices are motivated by the three circuit choices discussed in  \cite{Caputa:2022yju}. Here, in our work we use the reference-state choices only, rather than the spread-complexity construction itself. The construction in Ref.~\cite{Caputa:2022yju} uses antiperiodic
boundary conditions for even system size, whereas the present
reference-state analysis is performed with open boundary conditions. The spread-complexity plateaux reported there therefore do not imply
identical behavior of the spatial pseudo-entropy considered here.
\\

In the topological regime, a long Kitaev chain has two end Majorana fermions that combine to form a low-energy nonlocal fermionic mode \cite{Kitaev:2000nmw}. The empty and occupied states have opposite fermion parity,  while their energy splitting
becomes exponentially small as the chain length increases. The normalized transition matrix for the pseudo-entropy contains the overlap of two states. If the two states have definite fermion parity,
\begin{equation}
 P|\Psi_s\rangle=p_s|\Psi_s\rangle,
 \qquad p_s=\pm1,
\end{equation}
then, using $P^2=I$, their overlap obeys
\begin{equation}
 \langle\Psi_2|\Psi_1\rangle
 =p_1p_2\langle\Psi_2|\Psi_1\rangle.
 \label{eq:refdep-parity-overlap}
\end{equation}
For opposite parities, $p_1p_2=-1$, this relation implies
$\langle\Psi_2|\Psi_1\rangle=0$. Consequently, the normalized
transition matrix is not defined for a pair of states with opposite
fermion parity. We therefore use the same $|\Psi_+(\mu)\rangle$ of
Eq.~\eqref{eq:refdep-target} in all comparisons and select every
reference in the even-parity sector. The pairing-only and vacuum
references already have even parity. For the free-fermion reference,
we take the lowest-energy state within the even sector, which need
not be the absolute free-fermion ground state. Thus the parity
of the target does not change when the reference changes. At
an exact free-reference filling degeneracy, the selected state is
not unique, so we report one-sided limits rather than choosing an
arbitrary state. Equal parity is necessary, but does not by itself guarantee a nonzero overlap.  \\

\begin{figure*}[t]
    \centering
    \includegraphics[width=1\textwidth]{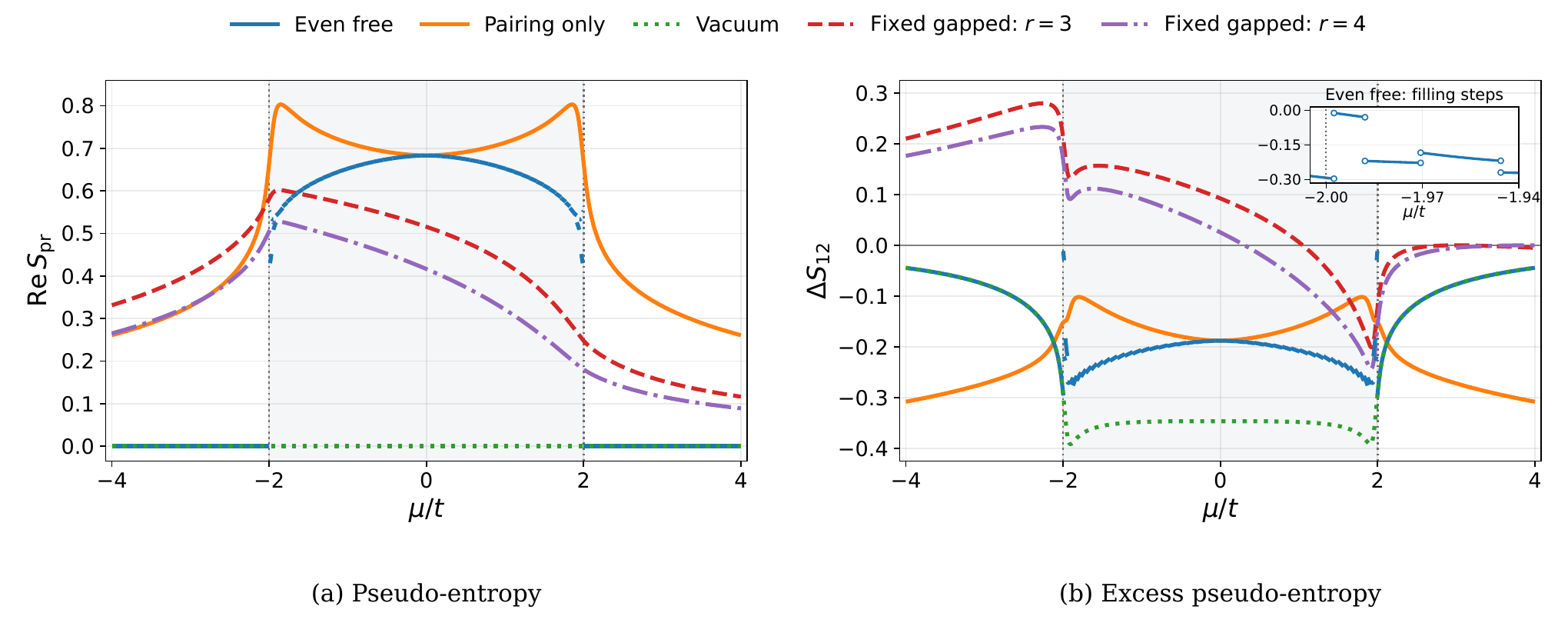}
     \caption{ Reference-state dependence of the pseudo entropy for the open Kitaev chain with \(L=100\), \(\ell=50\), and target parameters \(t=\Delta=1\). The same even-parity target is used in all comparisons and  this plot represents the even-$\ell$ sequence. (a) Real part of pseudo entropy and (b) averaged excess pseudo entropy \(\Delta S_{12}\) as functions of \(\mu/t\), for the even free-fermion, pairing-only, vacuum, and two fixed gapped references with \(\mu_{\mathrm{ref}}/t=3,4\). The shaded region denotes the topological interval \(|\mu|<2t\), and the vertical dotted lines mark the bulk transitions. The inset in (b) shows the free-reference filling steps near \(\mu/t=-2\), with separate segments at the filling degeneracies. 
}
 \label{fig:refstatecombined}
\end{figure*}

The parity selection must also be implemented in the Gaussian
state used to construct the transition correlator. Starting from the negative-energy occupied Nambu frame, we
change the occupation of the lowest quasiparticle when needed by
replacing an occupied vector with its particle--hole partner,
$u_a\mapsto\tau_xu_a^*$. For a numerically unresolved edge
zero mode, the occupation is assigned explicitly by parity. The
canonical frame conditions were given in the preceding subsection and the occupation selection is checked against the parity obtained from the Gaussian covariance.
 We therefore verify that the overlap matrix
$U_2^\dagger U_1$ is nonsingular. Its singular values determine the
many-body overlap through the Onishi relation \cite{onishi1966generator,Robledo:2009yd,Ring:1980}
\begin{equation}
 \log\left|\langle\Psi_2|\Psi_1\rangle\right|
 =\frac{1}{2}\sum_{a=1}^{L}
 \log\sigma_a\!\left(U_2^\dagger U_1\right).
 \label{eq:iv-overlap}
\end{equation}
Here, the smallest singular value measures proximity to a singular
transition correlator, whereas the full set of singular values
determines the total many-body overlap. Eqn.~\ref{eq:iv-overlap} fixes only the modulus of the overlap, and its sign is
obtained from the Pfaffian overlap formula~\cite{Bertsch:2011pr},
which is not required to construct the phase-invariant normalized Gaussian correlator. The transition matrix itself is normalized by the full complex overlap, not only its modulus. We use a linear solve rather than a pseudoinverse, and reject singular comparisons.
For the five reference choices considered here, the overlap is nonzero at finite even $L$ away from free-reference filling degeneracies. In the common occupation basis, the even-parity target ground state has strictly positive amplitudes for real finite $\mu$, while each selected reference can be chosen with nonnegative amplitudes on a nonempty subset of the same even sector. Their overlap is therefore strictly positive with this phase convention. This statement concerns the specified reference states, rather than arbitrary even-parity states. A nonzero overlap can nevertheless be exponentially small, and numerical conditioning must be checked separately. 
For the five references in
Fig.~\ref{fig:refstatecombined}, the accepted
$L=100$ mode weights are real within numerical
precision (the largest sampled imaginary component
is below $1.1\times10^{-14}$). Only the even
free-fermion reference has resolved negative mode
factors, giving a maximum sampled
$|\operatorname{Im}S_{\mathrm{pr}}|\simeq0.1148$.
The other four references have imaginary parts
below $9\times10^{-14}$ in the same scan. These are numerical observations for the sampled
parameter range; the endpoint prescription still
allows a nonzero imaginary pseudo entropy for real
mode weights.

\subsubsection{Free-fermion reference}

In this  reference state, we keep the target's hopping and chemical potential
but remove the pairing. Its open-chain single-particle energies are
\begin{equation}
 \epsilon_m(\mu)=-\mu-2t\cos\frac{m\pi}{L+1},
 \qquad m=1,\ldots,L,
 \label{eq:iv-free-levels}
\end{equation}
with $t=1$. The lowest even-parity reference changes its filling by two
fermions as $\mu$ varies, while the reference parity remains even.
A transition between fillings $n$ and $n+2$ occurs at
\begin{align}
\mu_{n\to n+2}
={}&-\cos\frac{(n+1)\pi}{L+1}\nonumber\\
&-\cos\frac{(n+2)\pi}{L+1},\quad n=0,2,\ldots,L-2.
\label{eq:iv-even-filling}
\end{align}
These filling thresholds are distinct from those of the
unconstrained free-fermion sea. The last positive threshold
approaches the target bulk boundary with
\begin{equation*}
2-\mu_{\mathrm{fill}}=\frac{5\pi^2}{2(L+1)^2}+O(L^{-4}).
\end{equation*}
Below the first negative threshold the even free reference is exactly the vacuum and above the last positive threshold it is the fully occupied product state. The pseudo entropy therefore vanishes in these intervals whenever the transition operator is defined. The free-reference band edges at $|\mu|=2t$ coincide with the target bulk boundaries, so this reference alone cannot separate reference-driven and target-driven features.\\

Figure~\ref{fig:refstatecombined}(a) shows a nonzero pseudo entropy
inside the topological region for the even free-fermion reference.
Its finite-size structure is affected by the discrete filling
changes of the reference. In particular, close to $\mu=2$ the even reference becomes a fully occupied product state in a narrow interval that still lies within the bulk topological region, and its pseudo entropy vanishes there. At exact filling thresholds the reference is degenerate and the plotted curve shows one-sided segments rather than an interpolated value. Therefore, not every jump or
feature in this curve can be interpreted as a transition of the
target state. This further shows that the response depends on both
the target and the reference state.

\subsubsection{Pairing-only reference}

In this  reference state, we make the pairing-only reference independent of $\mu$. The hopping term is 0 while the pairing term is finite. For even $L$, the ground state of this reference state is nondegenerate and the smallest
positive BdG energy is
\begin{equation}
 \epsilon_{\min}^{(\Delta)}
 =2|\Delta|\sin\frac{\pi}{2(L+1)}.
 \label{eq:iv-pairing-gap}
\end{equation}
This energy decreases as $1/L$, showing that the pairing-only
reference becomes gapless in the thermodynamic limit.

The pseudo entropy remains nonzero in both trivial regions
of the target and is enhanced near the bulk transitions, as shown
in Fig.~\ref{fig:refstatecombined}(a). The height of the boundary
peak changes with system size because this reference is critical
in the thermodynamic limit. 
This finite-size growth should not be interpreted as a universal
peak scaling law. At $\mu=0$, the central value close to $\ln2$ follows distinct exact finite-size results for $L=4m$ and $L=4m+2$, as discussed in the following subsection. A smaller even--odd subsystem alternation is also visible away from $\mu=0$ in the finite-size data.
Since the pairing-only reference does not change with $\mu$, the
variation of the pseudo entropy comes from the changing
target state. However, the magnitude and shape of the response still
depend on the correlations present in the chosen reference state.
Therefore, the value close to $\ln 2$ should not by itself be
interpreted as a universal contribution from the Majorana modes.

\subsubsection{Vacuum reference}

The vacuum reference state provides an exact limit of reference dependence. This follows from the known product-reference property~\cite{Nakata2021} and is not a new topological result. For any product reference
$|\pi\rangle=|\pi_A\rangle\otimes|\pi_{\bar A}\rangle$
with $\langle\Psi|\pi\rangle\ne0$, define
$\langle\chi_A|=\langle\Psi|(I_A\otimes|\pi_{\bar A}\rangle)$.
The reduced transition matrix is
\begin{equation}
 \mathcal T_A^{\pi|\Psi}
 =\frac{|\pi_A\rangle\langle\chi_A|}
        {\langle\Psi|\pi\rangle},
 \qquad
 (\mathcal T_A^{\pi|\Psi})^2=\mathcal T_A^{\pi|\Psi}.
 \label{eq:iv-product}
\end{equation}
This rank-one operator has unit trace and spectrum
$\{1,0,\ldots,0\}$, so its pseudo entropy vanishes.
For the vacuum reference,
\begin{equation}
 S_{\mathrm{pr}}(\mathcal T_A^{0|\Psi_+})=0,\qquad
 \Delta S_{12}=-\frac12S(\rho_A^{(\mu,+)}).
 \label{eq:iv-vacuum}
\end{equation}
The result holds independently of Gaussianity and of the
target phase, provided the overlap is nonzero.
Figure~\ref{fig:refstatecombined} confirms that the pseudo entropy vanishes for the vacuum reference, while the averaged excess pseudo entropy still contains the
entanglement of the target state.

\subsubsection{Fixed gapped references}

To examine whether the enhancement of pseudo entropy
requires a critical reference, we also consider two fixed,
gapped and entangled reference states,
\begin{equation}
|\Psi_r\rangle=\operatorname{GS}_{P=+1}[H(1,r,1)],
\qquad r=3,4.
\label{eq:iv-fixed-refs}
\end{equation}
Both are in the trivial Kitaev phase and remain gapped
as $L$ increases. Their lowest positive quasiparticle energy satisfies $\epsilon_{\min}\ge r-2$ by an open-chain singular-value bound, and their spatial entanglement is non-zero. In the even sector, the overlap with the target is
strictly positive at finite $L$. We keep these reference states
fixed while changing $\mu$ in the target.

Figure~\ref{fig:refstatecombined}(a) shows that the fixed gapped references respond near the target bulk boundaries, but the two sides are not symmetric. For the even-$\ell$ sequence $L=100,200,400,800$, the interior maxima of $\operatorname{Re}S_{\mathrm{pr}}$ near $\mu/t=-2$, on the topological side, change only weakly with size. At $L=800$ these maxima are approximately $0.6073$ and $0.5345$ for $r=3$ and $r=4$, respectively. This behavior is consistent with saturation over the tested even-$\ell$ sizes, but does not establish a thermodynamic limit. In the corresponding windows near $\mu/t=+2$, no interior maximum was found.

The odd-$\ell$ sequence exhibits a different response, particularly for $\mu/t<-2$. At $\mu/t=-4$, $\operatorname{Re}S_{\mathrm{pr}}$ changes from $0.33125$ to $0.68001$ for $r=3$, and from $0.26505$ to $0.69375$ for $r=4$, when the chain length is changed from $L=100$ ($\ell=50$) to $L=102$ ($\ell=51$). The distinction persists across the tested larger sizes and changes the shape of the negative-boundary response, not only its height. Thus the curves in Fig.~\ref{fig:refstatecombined}, which use even $\ell$, should not be taken as representative of both subsystem-parity sequences. In the opposite-field case $\mu=-r$ with odd
$\ell$, the paired positive spectrum gives
$S_{\mathrm{pr}}\ge\ln2$. This bound is specific to that setting and is not a general property of the Majorana phase.

The excess pseudo entropy is also positive in part of the region $\mu<-2$, where the target and both fixed references have trivial values of the bulk Majorana $\mathbb{Z}_2$ invariant. The target is vacuum-like in this region, whereas the positive-$r$ references are filled-like, and their many-body overlap can be exponentially small even though it is nonzero at finite size. Thus positive averaged excess does not require a Majorana-topological target, and its sign cannot be inferred from the target invariant alone. This finite-size observation does not by itself establish a general counterexample to the MNT conjecture. The fixed-reference curves have no free-fermion filling jumps, and their shapes depend on the reference correlations. For the real, reflection-symmetric Hamiltonians and contiguous equal-half partition considered here, the reduced-transition spectrum of these fixed gapped comparisons is nonnegative; the statement follows from the positive-semidefinite half-chain coefficient matrices.
Together, the different reference choices show that the
pseudo entropy is not determined solely by the target phase. The
free reference carries filling-dependent structure, the pairing-only
reference is critical in the thermodynamic limit, and the vacuum
gives zero pseudo entropy even for an entangled target. The fixed
gapped references provide an additional control: they can respond
near the target bulk transitions without themselves becoming
critical. The excess can also be positive when the target is in a trivial region. The different subsystem-parity sequences provide another example of the dependence on the reference--target pair and partition. We therefore interpret the pseudo-entropy response as a property of the reference--target pair and subsystem geometry, rather than a reference-independent topological invariant.\\

\subsection{Central pseudo entropy and finite-size behavior}
\label{sec:iv-central}

We now examine the central value of the pseudo entropy for the
pairing-only reference in more detail. In the previous subsection,
we found that this reference gives a value close to $\ln 2$ at
$\mu=0$. We show here that this behavior depends on the parity of
the subsystem size, and that its finite-size correction can be
obtained analytically. This provides an example of how the reference
state and the symmetry of the target together determine the pseudo
entropy.

For $t=\Delta=1$ and $\mu=0$, the even-parity target state has the
Schmidt form
\begin{equation}
 |\Psi_+(0)\rangle
 =\frac{|E_AE_{\bar A}\rangle+|O_AO_{\bar A}\rangle}{\sqrt{2}},
 \label{eq:iv-central-schmidt}
\end{equation}
where $|E\rangle$ and $|O\rangle$ are normalized sums of the states
with even and odd local fermion parity, respectively. Thus the target
has Schmidt rank two. For the pairing-only reference $|\Psi_\Delta\rangle$,
the reduced transition matrix has at most two nonzero eigenvalues,
which can be written as
\begin{equation}
 q_\pm=\frac{1\pm\nu_L}{2},\qquad
 \nu_L=\frac{\langle\Psi_+(0)|P_A|\Psi_\Delta\rangle}
 {\langle\Psi_+(0)|\Psi_\Delta\rangle},
 \label{eq:iv-central-weights}
\end{equation}
with $P_A=(-1)^{\sum_{j\in A}n_j}$. The overlap in the denominator
is nonzero for every finite even $L$.

At $\mu=0$, both states are invariant under the same particle--hole
symmetry. This symmetry changes the sign of $P_A$ when
$\ell=L/2$ is odd. As a result, $\nu_L=0$ for $L=4m+2$, so that
$q_+=q_-=1/2$. We obtain the exact finite-size result
\begin{equation}
 S_{\mathrm{pr}}(0,L)=\ln 2,\qquad L=4m+2.
 \label{eq:iv-central-odd}
\end{equation}
This equality holds at every finite size in this sequence. It is not
only a large-$L$ limit.

For $L=4m$, the subsystem size is even and the same symmetry does
not require $\nu_L$ to vanish. In this case, an exact calculation
gives
\begin{equation}
 \nu_L=\prod_{j=1}^{m}
 \frac{\sin[(2j-1)\pi/(L+1)]}
 {\sin[2j\pi/(L+1)]},\qquad L=4m,
 \label{eq:iv-central-product}
\end{equation}
with $q_\pm$ given by Eq.~\eqref{eq:iv-central-weights}. The pseudo
entropy is therefore
\begin{equation}
 S_{\mathrm{pr}}(0,L)
 =-q_+\ln q_+-q_-\ln q_-.
 \label{eq:iv-central-binary}
\end{equation}
Both weights are positive, and the central pseudo entropy is real.
The product formula follows from the pairing-reference overlap and
a Pfaffian reduction.

\begin{figure*}[t]
    \centering
    \includegraphics[width=1\textwidth]{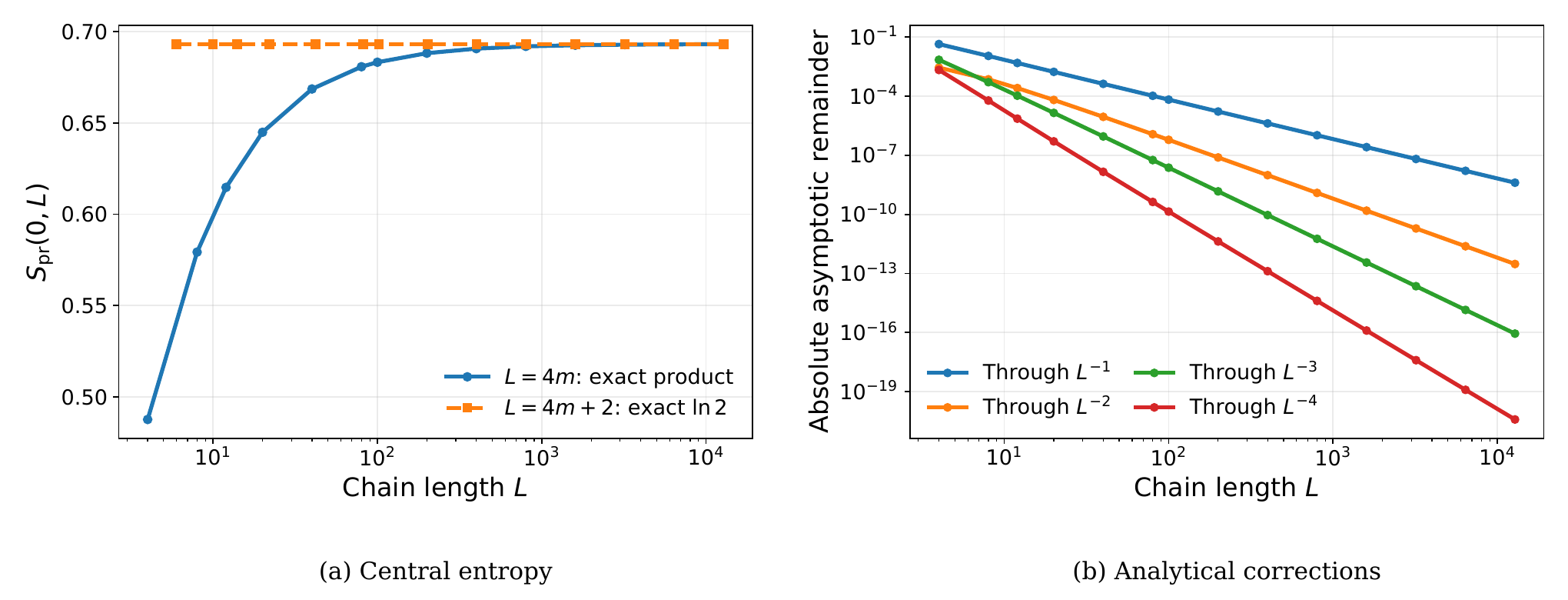}
     \caption{Central pseudo entropy for the pairing-only reference
at $\mu=0$ in the open Kitaev chain with $t=\Delta=1$
and $\ell=L/2$.
(a) Exact finite-size values for $L=4m$ obtained from the
finite sine product, compared with the exact result
$S_{\mathrm{pr}}=\ln 2$ for $L=4m+2$.
(b) Absolute remainders of the entropy deficit
$d_L=\ln 2-S_{\mathrm{pr}}(0,L)$ for $L=4m$,
after retaining terms through orders $L^{-1}$,
$L^{-2}$, $L^{-3}$, and $L^{-4}$ in the analytical
expansion. The coefficients are obtained analytically
without numerical fitting.}
 \label{fig:central_result}
\end{figure*}

For $L=4m$, the central comparison with the even free-fermion
reference is also exactly related to the pairing-only result.
The on-site transformation $R=\prod_{j\,\mathrm{even}}X_j$
maps $|\Psi_\Delta\rangle$ to the half-filled free-fermion
ground state at $\mu=0$ and leaves $|\Psi_+(0)\rangle$
unchanged. Since $R=R_A\otimes R_{\bar A}$, the reduced
transition matrices are related by a unitary transformation on
$A$. Thus their spectra, pseudo entropies and averaged excess
pseudo entropies agree. For $L=4m+2$, the half-filled free state
has odd total parity, and the lowest even free-reference state
is degenerate at $\mu=0$. The same equivalence therefore does
not apply to the parity-selected free reference in that sequence.

Writing $d_L=\ln2-S_{\mathrm{pr}}(0,L)$ and using this exact
expression, the finite-size correction for $L=4m$ is
\begin{equation}
 \ln 2-S_{\mathrm{pr}}(0,L)
 =\frac{1}{L}-\frac{2}{3L^2}+O(L^{-3}).
 \label{eq:iv-central-expansion}
\end{equation}

Figure~\ref{fig:central_result}(a) shows the two
subsystem-parity sequences of the central pseudo
entropy at $\mu=0$. Figure~\ref{fig:central_result}(b)
compares the exact finite-size result with the
analytical expansion for $L=4m$, showing how the
corrections decrease as more terms are included.
The two leading coefficients are obtained analytically, rather than
from a numerical fit. The correction does not apply to
$L=4m+2$, where Eq.~\eqref{eq:iv-central-odd} holds exactly.

The result at $\mu=0$ follows from the rank-two structure of the
target, but the values of the two transition eigenvalues still depend
on the reference. For example, the fixed gapped references considered
in the previous subsection give central pseudo entropies close to
$0.51571$ and $0.41650$ for $r=3$ and $r=4$ at the largest tested
sizes. These are numerical values, not exact limiting constants.
The exact $\ln 2$ result for $L=4m+2$ follows from the
symmetry of the pairing-only comparison; it is not a universal
value for the Majorana topological phase. For the pairing-only
reference at $\mu=0$, the two positive central weights imply
$S_{\mathrm{pr}}(0,L)\leq\ln 2$. This bound need not hold for
other reference--target pairs or away from $\mu=0$.

Finally, the pseudo entropy and the averaged excess pseudo entropy
do not have the same finite-size behavior. The pairing-only reference
is equivalent, up to a local unitary transformation, to a half-filled
open hopping chain. Its ordinary entanglement entropy grows as
$S_{\mathrm{ref}}=\frac{1}{6}\ln L+O(1)$~\cite{Calabrese:2004eu}.
Since $S_{\mathrm{target}}(0)=\ln 2$, we obtain
\begin{align}
 \Delta S_{12}(0,L)
 &=\tfrac12\ln 2-\tfrac12S_{\mathrm{ref}}(L)-d_L,\nonumber\\
 &=-\tfrac1{12}\ln L+O(1).
 \label{eq:iv-central-excess}
\end{align}
The logarithmic term comes from the critical reference state.
Therefore, even when the central pseudo entropy is exactly $\ln 2$
or approaches it with increasing system size, its excess need not
approach a constant. This further shows why the reference state
must be taken into account when interpreting pseudo entropy.

\section{Conclusion and Outlook}
\label{sec:conclusion}

In this work, our central question was how pseudo-entropy and the averaged excess entropy $\Delta S_{12}$ in the Rice--Mele (RM) model under Thouless pumping and in the Kitaev chain characterize topological features and depend on the choice of reference state. For the one-dimensional Rice--Mele model, we find that a nonzero staggered potential $\mu_s$ breaks chiral symmetry and opens a gap at the SSH critical point $|v_2|=w$. As $\mu_s$ increases, the peak of the pseudo-entropy becomes lower and broader, so the sharp signature of the phase boundary disappears. Along the pumping cycle, $\Delta S_{12}$ distinguishes the trivial, monomer, and topological stages of the control-freak protocol, and it depends on the boundary condition, the protocol (in our work, the control-freak limit and smooth pumping), and the reference time. In particular, the smooth pump under PBC shows sharp spikes at $t/T=t_{\rm ref}\pm0.5$, which are absent under OBC. Because these spikes move with $t_{\rm ref}$, they reflect the reference--target pair and not the instantaneous Hamiltonian alone. Reference states in the non-trivial and trivial SSH phases also give distinct $\Delta S_{12}$ curves, so the reference must be specified whenever $\Delta S_{12}$ is used to probe the pumping cycle.

For the open Kitaev chain, we compared five reference families with the same even-parity superconducting target. The parity-constrained free-fermion reference changes its filling in steps of two as $\mu$ varies, producing features that are not transitions of the target state. The pairing-only reference is critical in the thermodynamic limit, while the two fixed gapped references still respond near the target bulk boundaries and show a strong dependence on subsystem parity. The vacuum reference provides the known product-state limit of vanishing pseudo entropy whenever the overlap is nonzero. In the pairing-only comparison at $\mu=0$, we obtained the exact result $S_{\mathrm{pr}}=\ln 2$ for $L=4m+2$. For $L=4m$, the central entropy approaches this value with the analytically derived deficit $\ln 2-S_{\mathrm{pr}}(0,L)=L^{-1}-2/(3L^2)+O(L^{-3})$. The fixed gapped comparisons also show that positive excess pseudo entropy can occur when the target lies in a Majorana-trivial region. Taken together, these results show that the pseudo-entropy response depends on the reference--target pair and subsystem partition, and that a feature near $\ln 2$ or a positive excess cannot by itself be regarded as a reference-independent signature of Majorana topology.

Several interesting future directions remain open. One natural extension
is to study whether the pseudo-entropy features observed during Thouless
pumping remain robust for other pumping
protocols~\cite{Citro:2022cwl,Nakajima:2016zft,Lohse:2015qem} and
different choices of reference states. It would also be interesting to
examine the behavior of pseudo-entropy in the presence of
disorder~\cite{Motrunich:2001,Brouwer:2011,DeGottardi:2012ygr,Wauters:2019zgb}
and interactions~\cite{Fidkowski:2009dba,Fidkowski:2010jmn,Stoudenmire:2011,Katsura:2015yda},
where the correlation-matrix methods used in this work may no longer be
directly applicable. For the Kitaev chain, a further direction is to
identify classes of reference states and subsystem partitions that provide
more robust signatures of the Majorana topological phase. Finally,
comparing pseudo entropy with other quantities, such as the entanglement
spectrum~\cite{Li:2008kda,Pollmann:2009ryx,Fidkowski:2010nhf,10461354},
topological invariants~\cite{Zak:1989wzd,Budich:2013qea,adhikari2026certifiedfidelitysusceptibilityclassical}, and spread
complexity~\cite{Balasubramanian:2022tpr,Caputa:2022eye,t1bf-fxb9,Adhikari:2022whf,Adhikari:2025vdl,Adhikari_2024}, may help
clarify what additional information pseudo entropy provides about
topological systems.

\acknowledgments

We acknowledge the support from Abdus Salam International Centre for Theoretical Physics (ICTP) program: Physics Without Frontiers (PWF). 

\bibliographystyle{apsrev4-2}
\bibliography{refRM.bib}

\clearpage
\onecolumngrid

\appendix

\end{document}